\documentclass[seceq,preprint]{ptptex}

\usepackage{bm}

\newcommand{\wt}{\widetilde}
\newcommand{\wh}{\widehat}
\newcommand{\ol}{\overline}
\newcommand{\del}{\partial}
\newcommand{\ra}{\rightarrow}

\newcommand{\lra}{\leftrightarrow}
\newcommand{\nn}{\nonumber}
\newcommand{\half}{\frac{1}{2}}

\def\tr{\mathop{\rm tr}\nolimits}
\def\Tr{\mathop{\rm tr}\nolimits}

\newcommand{\vev}[1]{\left\langle #1 \right\rangle}

\newcommand{\cF}{{\mathcal F}}

\newcommand{\cO}{{\mathcal O}}
\newcommand{\cA}{{\mathcal A}}

\newcommand{\cM}{{\mathcal M}}
\newcommand{\bR}{\mathbb{R}}

\newcommand{\bH}{\mathbb{H}}
\newcommand{\bZ}{\mathbb{Z}}

\newcommand{\ba}{{\bm{a}}}
\newcommand{\br}{{\bm{r}}}
\newcommand{\bu}{{\bm{u}}}
\newcommand{\bq}{{\bm{q}}}
\newcommand{\bw}{{\bm{w}}}
\newcommand{\by}{{\bm{y}}}

\newcommand{\bx}{{\bm{x}}}
\newcommand{\bX}{{\bm{X}}}

\newcommand{\Mkk}{M_{\rm KK}}

\newcommand{\vk}{\vec{k}}
\newcommand{\vp}{\vec{p}}
\newcommand{\vx}{\vec{x}}

\def\mat#1{\matt[#1]}
\def\matt[#1,#2,#3,#4]{\left(%
\begin{array}{cc} #1 & #2 \\ #3 & #4 \end{array} \right)}

\notypesetlogo                       
\preprintnumber[3cm]{
IU-MSTP-77\\IPMU 09-0010\\ RIKEN-TH-147}

\title{
Nuclear Force from String Theory
}

\author{
Koji \textsc{Hashimoto}$^1$,
Tadakatsu \textsc{Sakai}$^2$, and Shigeki \textsc{Sugimoto}$^3$
}

\inst{$^{1}$Theoretical Physics Laboratory, RIKEN, 
Wako 351-0198, Japan \\ 
$^{2}$Department of Physics, Faculty of Sciences, Ibaraki University, \\
Mito 310-8512, Japan \\ 
$^{3}$Institute for the Physics and Mathematics of the Universe,\\
The University of Tokyo, Chiba 277-8568, Japan}

\abst{
We compute the nuclear force in a holographic model of QCD 
on the basis of a D4-D8 brane configuration in type IIA string theory.
The repulsive core of nucleons is important
in nuclear physics, but its origin has not been
well understood in strongly coupled QCD.
We find that the string theory via gauge/string duality deduces 
this repulsive core at a short distance
between nucleons. Since baryons in the model are realized as
solitons given by Yang-Mills instanton configuration
on flavor D8-branes, ADHM construction of two
instantons probes well the nucleon interaction at short scale,
which provides the nuclear force quantitatively. We obtain a central
force, as well as a tensor force, which is strongly repulsive as suggested
in experiments and lattice results. 
In particular, the nucleon-nucleon potential $V(r)$ (as a function of the
distance) scales as $r^{-2}$, which is peculiar to the holographic
model.
We compare our results with the one-boson
exchange model using the nucleon-nucleon-meson coupling obtained in our
previous paper \cite{HSS}.
}

\begin{document}

\maketitle

\section{Introduction}
\label{sec:1}

Nuclear force, the force between nucleons, exhibits a repulsive core of
nucleons at short distances. This repulsive core is quite important for 
large varieties of 
physics of nuclei and nuclear matter. For example,
the well-known presence of nuclear saturation density is essentially 
due to this repulsive core. However, 
from the viewpoint of strongly coupled QCD,
the physical origin of this repulsive
core has not been well understood. 
Despite the long history of the problem, it was rather recent 
\cite{Ishii:2006ec} that lattice QCD could reach the problem,\footnote{
See also Ref.~\citen{Beane} for a study of the
interactions between nucleons and hyperons using lattice QCD.}
and of course, any understanding
of it based on analytic computations is quite helpful for revealing 
the basic nature of nuclear and hadron physics.
\footnote{For a review on the theoretical aspects of nuclear force including
that of short distance, see for example Ref.~\citen{Myhrer:1987af}. }

The recent rapid progress in applying gauge/string duality 
\cite{Maldacena:1997re,Gubser:1998bc,Witten:1998qj,Aharony:1999ti}
to QCD, holographic QCD, has been
surprising. Now, it has been made possible to compute various observables in hadron
physics such as spectra of mesons/baryons/glueballs and the interactions
among them.
Although most of the works rely on the supergravity approximation
that works for large $N_c$ and large 't Hooft coupling $\lambda$,
it turned out that the holographic QCD
reproduces quite well the properties of hadrons
not only qualitatively but also quantitatively. 

We apply this gauge/string duality to the problem of nuclear force. 
In our previous paper \cite{HSS},\footnote{
See also Refs.~\citen{Hong-Rho-Yee-Yi,Hata-Murata-Yamato,
Kim:2008pw,Panico:2008it,Seki:2008mu} for closely related works.
}
 we computed nucleon-nucleon-meson
couplings, using the holographic QCD on the basis of a
 D4-D8 brane configuration in type IIA string theory \cite{SaSu1,SaSu2},
which incorporates chiral quark dynamics.
This amounts in principle to
computing the large distance behavior of nuclear force, given that the
potential between two nucleons can be understood as an exchange of
mesons between them.
In this paper, we take one step further: by directly solving the 
two-nucleon system in the D4-D8 model of the holographic QCD, we find
a short distance scale of the nuclear force. In fact, we find the
repulsive core of nucleons.

First, let us briefly summarize what has been computed for baryons
in the D4-D8 model of the holographic QCD. 
The D4-D8 model \cite{SaSu1,SaSu2} of the
holographic QCD describes a strong coupling regime of massless QCD at
low energy, in the large $N_c$ limit with large 't Hooft coupling
$\lambda$, for a fixed number $N_f$ of flavors.\footnote{Introducing
massive quarks in the model has been discussed in 
Ref.~\citen{Aharony:2008an}.}
 The low-energy degrees
of freedom on the flavor D8-branes in the holographic geometry 
of Ref.~\citen{Witten:D4}, which are basically the Yang-Mills (YM)
fields in five-dimensional curved space-time, give Kaluza-Klein towers
of mesons, while instantons in the YM theory correspond to
baryons in low-energy QCD \cite{SaSu1} (this is based on the baryon
vertices in gauge/string duality \cite{Gross-Ooguri, WittenBaryon} 
and the fact that branes
inside branes are represented by solitonic instantons
\cite{brane-within-brane}).
Here, in our terminology, the instanton is a gauge configuration that is
localized in spatial four dimensions in the five-dimensional space-time.
The baryon number is identified as the instanton number in
four-dimensional space. Since it is conserved in the time direction
and localized in the spatial directions, it behaves as point particles
that are interpreted as baryons.
Quantization of a single instanton {\it a la} moduli space
approximation \cite{GeSa,Manton}
gives rise to a spectrum of baryons including
nucleons \cite{HSSY}. In our previous paper \cite{HSS}, 
we computed the static properties of the
baryons by evaluating the chiral currents in the presence of the
instanton: charge radius, magnetic moments, form factors etc.
were computed, in
addition to the nucleon-nucleon-meson couplings.\footnote{
By using five-dimensional spinor fields introduced as nucleon fields 
on the D8-brane, in Refs.~\citen{Hong-Rho-Yee-Yi,PaYi},
the static quantities of baryons were computed.
See also Refs.~\citen{Hata-Murata-Yamato,Kim:2008pw}.}
This analysis is reminiscent of that by Adkins et. al.\cite{ANW}
for Skyrmions \cite{Skyrme}.\footnote{
The analysis of the Skyrmions based on the four-dimensional
meson effective action derived from the D4-D8 model is given in
Ref.~\citen{NaSuKo}.}
In fact, the 
holographic description mimics the relation between the Skyrmion and
instantons found by Atiyah and Manton \cite{Atiyah-Manton}\footnote{
To describe the nuclear force, a two-instanton configuration was used for this
Atiyah-Manton ansatz for Skyrmions (see for example Ref.~\citen{2is}).
}.
The physics of finite baryon density and nuclear matter 
has been explored in many papers recently, and we do not describe
them in detail here.

Next, we briefly outline our method. The one-instanton analysis given in
Refs.~\citen{HSSY,Hong-Rho-Yee-Yi} 
revealed that the desired configuration with
instanton number 1 can be obtained simply by considering corrections to
the BPST instanton in four-dimensional flat space \cite{BPST}.
The corrections are due to (i) overall $U(1)$ part of the YM gauge fields 
coupled to the instanton density, and (ii) curved space-time along the
extra dimension $x^4$ in the five-dimensional space-time.
These corrections induce a small potential in the instanton moduli
space, fix the size of the instanton to be of order
$1/(\sqrt{\lambda}M_{\rm KK})$ (where $M_{\rm KK}$ is the only parameter with
mass dimension and gives the meson mass scale), and give the quantization
of the instanton in the moduli space approximation. This type of analysis
can be extended to our case of two baryons. If the two baryons sit close to
each other so that the distance $r$ satisfies 
$r < {\cal O}(1/M_{\rm KK})$, we can use two-instanton configuration in
the flat space as a starting point, since the effects of the curved space
are small. The properties of the 
two-instanton moduli space are known, 
concerning not only its construction via renowned ADHM 
(Atiyah-Drinfeld-Hitchin-Manin) method \cite{ADHM}, 
but also the metric in its moduli space (see
Refs.~\citen{ADHMreview}--\citen{PeZa} for some of the papers relevant to
our computations).
We use them explicitly as a
basis in a manner similar to the one-instanton case, to explore the physics
of the nuclear force, {\it i.e.}, the interaction between two baryons
sitting close to each other.

We compute the additional potential induced in the moduli space, due to
the presence of the two instantons. The analytic form of the moduli
Lagrangian can be obtained in the asymptotic expansion of $r$. This
includes corrections to the kinetic term, coming from the metric of
two-instanton moduli space. Specifying the two-nucleon states by
tensor product of single-baryon states obtained in Ref.~\citen{HSSY},
we can evaluate the nuclear force for given nucleon states.

Note that
this ``asymptotics'' means a large distance in the region 
$r<{\cal O}(1/M_{\rm KK})$. Therefore, in the standard terminology for
the nuclear force, our result is for short distances. 
In addition, we use the asymptotic expansion in $r$, so 
our analytic formula of the nucleon-nucleon potential is not for 
nucleons on top of each other. However,
this is sufficient for seeing the repulsive core of the nucleons.

We find that our final expression for the nucleon-nucleon potential,
(\ref{fresult0}), 
is repulsive, and has $1/r^2$ dependence. This
$r$-dependence is peculiar to the four-dimensional space, not the
three-dimensional harmonic potential. The appearance of
the $1/r^2$ potential is due to the extra holographic dimension, thus
typical in holographic description. Physically speaking, the
Kaluza-Klein summation of all the meson states in the tower 
produces this new behavior.

The main reason why the force is repulsive is that the instantons carry
electric charge of the overall $U(1)$ part of the YM fields on
the D8-branes. This electric charge is supplied by a Chern-Simons (CS)
coupling on the $N_f$ D8-branes, and is nothing but the baryon number.
The $U(1)$ force is repulsive since the 
instantons have the $U(1)$ charge of the same sign.
There are some contributions from the $SU(2)$ gauge field
that give attractive potential, but this $SU(2)$ force
turns out not to be strong enough to cancel the
$U(1)$ repulsive force. The Kaluza-Klein decomposition of the 
$U(1)$ part of the gauge fields provides a mass 
tower starting with $\omega$ meson as the lightest vector meson
\cite{SaSu1}, and so,
our computation shows that 
the repulsive force is partly due to the $\omega$ meson exchange.
Not only the $\omega$ meson 
but also the whole massive mesons participate in the nuclear 
force, and as a result, the nucleon-nucleon potential becomes $1/r^2$.

One might wonder whether it is reasonable to sum up the contributions
from all the massive mesons, since the model deviates from QCD
at the energy scale higher than $M_{\rm KK}$. However, there are some
lines of evidence suggesting that the results obtained by summing up the
infinite tower of massive mesons are better than those obtained
by only taking into account the first few modes.
For example, in our previous paper \cite{HSS}, we showed that the
electromagnetic form factors for the nucleon are very close to the dipole
profile observed in the experiment.
This result is obtained by summing up the contributions from all
the massive vector mesons. If we only take into account the rho meson,
the form factors can never be close to the dipole profile.

The organization of this paper is as follows. First, in
\S\ref{sec:2}, we describe our strategy, together with a brief review
of the instantons in the D4-D8 model. In \S\ref{sec:3}, 
we obtain an effective Hamiltonian for moduli
parameters of the two instantons in the model. In \S\ref{sec:4}, 
using the wave
functions for nucleon states, we evaluate the nucleon nucleon
interaction potential. We decompose it to a central force and a tensor
force. In \S\ref{sec:5}, we compare our results with
one-boson-exchange potential evaluated using the nucleon-nucleon-meson
coupling obtained in our previous paper \cite{HSS}. Section~\ref{sec:6} 
is for a brief summary. In the appendices, we review the ADHM
construction of two instantons and summarize the necessary formulas used in
this paper.

\section{Nuclear force in holographic QCD}
\label{sec:2}

In this section, we briefly summarize the treatment of the single
baryon in the D4-D8 model \cite{SaSu1,SaSu2}
of the holographic QCD following Ref.~\citen{HSSY} and
describe our strategy for obtaining the nuclear force.
Our first goal is to obtain a quantum mechanics Hamiltonian for 
a two-nucleon system.
The total Hamiltonian consists of
one-body canonical kinetic terms for each nucleon, potential terms
for each nucleon, plus interactions. One generically has
an interaction potential as well as a correction to the kinetic term.
Then, secondly, we evaluate the inter-baryon energy using the
Hamiltonian. This provides an explicit nuclear force that is 
dependent on nucleon states labeled by spin and isospin.

The concrete calculations of the Hamiltonian 
will be given in \S\ref{sec:3}, 
and its evaluation with explicit nucleon
states will be presented in detail in \S\ref{sec:4}.

\subsection{Review: single baryon in the model}
\label{sec:2-1}
First, we review briefly the single baryon case \cite{HSSY} in the
holographic QCD proposed in Refs.~\citen{SaSu1,SaSu2}.
The notation of our paper follows that of Ref.~\citen{HSSY}. 

Our starting point is the meson effective action derived in 
 Refs.~\citen{SaSu1,SaSu2}, which is given by
the following five-dimensional $U(N_f)$ Yang-Mills-Chern-Simons
(YMCS) theory in a curved
background:
\begin{align}
&S=S_{\rm YM}+S_{\rm CS}\ ,\nn\\
&S_{\rm YM}=-\kappa
\int d^4 x dz\,\tr\left[\,
\half\,h(z){\cF}_{\mu\nu}^2+k(z){\cF}_{\mu z}^2
\right]\ ,~~
S_{\rm CS}=\frac{N_c}{24\pi^2}
\int_{M^4\times\bR}\omega_5({\cA})\ .
\label{model}
\end{align}
Here $\mu,\nu=0,1,2,3$ are four-dimensional Lorentz indices, and $z$
is the coordinate of the fifth dimension. 
The field strength 
is defined as $\cF=\frac{1}{2}\cF_{\alpha\beta}
dx^\alpha\wedge dx^\beta=d\cA+i\cA\wedge\cA$ 
with the $U(N_f)$ gauge field
${\cA}=\cA_\alpha dx^\alpha=\cA_\mu dx^\mu+\cA_z dz
~~(\alpha=0,1,2,3,z)$, and the front factor $\kappa$ 
is related to  the 't~Hooft
coupling $\lambda$ and the number of colors $N_c$ as
\begin{equation}
\kappa=\frac{\lambda N_c}{216\pi^3}\equiv a\lambda N_c\ .
\label{kappa}
\end{equation}
The action \eqref{model} is written in the unit $M_{\rm KK}=1$,
where $M_{\rm KK}$ is the only dimensionful parameter
in the model.\footnote{
In Refs.~\citen{SaSu1} and \citen{SaSu2}, these two parameters
are chosen as
$\Mkk=949\mbox{ MeV},\kappa=0.00745$
to fit the experimental values of the $\rho$ meson mass 
$m_\rho\simeq 776\mbox{ MeV}$ and
the pion decay constant $f_\pi\simeq 92.4\mbox{ MeV}$.
}
The functions $h(z)$ and $k(z)$ appearing as the ``metric'' in the
action (\ref{model}) are given by $h(z)=(1+z^2)^{-1/3}$ and $k(z)=1+z^2$,
while, in the second term, $\omega_5(\cA)$ is the CS 5-form
(here, we omit the $\wedge$ 
product, e.g.,~$\cA\cF^2=\cA\wedge \cF\wedge\cF$)
\begin{equation}
\omega_5({\cA})=\tr\left(
\cA \cF^2-\frac{i}{2}\cA^3\cF-\frac{1}{10}\cA^5
\right)\ .
\end{equation}
In the two-flavor case ($N_f=2$) that we focus on in this paper,
the $U(2)$ gauge fields $\cA$  are decomposed as
\begin{eqnarray}
\cA=A+\wh A\,\frac{{\bf 1}_2}{2}
=A^a\frac{\tau^a}{2}+\wh A\,\frac{{\bf 1}_2}{2}
=\sum_{C=0}^3 \cA^C\,\frac{\tau^C}{2}\ ,
\label{decom}
\end{eqnarray}
where $\tau^a$ ($a=1,2,3$) are Pauli matrices and $\tau^0={\bf 1}_2$
is a unit matrix of size 2.

This action is obtained from the low-energy effective action on $N_f$
D8-branes in the curved ten-dimensional geometry corresponding to $N_c$
D4-branes wrapped on a circle with an antiperiodic boundary condition
for fermions.
At low energy, this D-brane configuration provides $U(N_c)$ QCD with
$N_f$ massless quarks and the action \eqref{model} describes the
dynamics of mesons and baryons.
The action (\ref{model}) is written in (1+4) dimensions, and the space
along the extra dimension $x^4(\equiv z)$ is curved. Once the gauge
fields are decomposed into their Kaluza-Klein states concerning the $z$
direction, each mass eigenstate corresponds to a meson, and the action
(\ref{model}) describes the whole spectra/interactions of the mesons.
By contrast, baryons are solitons with nonzero 
instanton number in the four-dimensional space parameterized by
$x^M=(\vec x,z)$ ($M=1,2,3,z$).
As they are localized in the four-dimensional space in the
five-dimensional space-time, they behave as pointlike particles.
The instanton number is identified with the baryon number \cite{SaSu1}
and these particles are interpreted as baryons.
We will see more of the details below.

The single-baryon solution was found to have the size of order 
$\lambda^{-1/2}$
\cite{Hong-Rho-Yee-Yi,HSSY}. It is helpful to rescale the coordinates
as \cite{HSSY}
\begin{eqnarray}
&& \widetilde{x}^M = \lambda^{1/2}x^M\ , 
\quad \widetilde{x}^0 = x^0\ , 
\nn\\
&&\widetilde{\cA}_0(t,\widetilde{x}) = 
\cA_0(t,\widetilde{x})\ , \quad
\widetilde{\cA}_M(t,\widetilde{x}) = \lambda^{-1/2}
\cA_M(t,\widetilde{x})\ ,
\label{rescalemod}
\end{eqnarray}
to see the consistent $1/\lambda$ expansion of the equations of
motion and the total energy of the single baryon. Hereafter, we omit the
tilde for simplicity. Then, for large $\lambda$, the 
YM part of the action is 
\begin{eqnarray}
S_{{\rm YM}}=&-aN_c\displaystyle\int d^4 x dz \,\tr\left[\,
\frac{\lambda}{2}\,F_{MN}^2+\left(
-\frac{z^2}{6} F_{ij}^2+z^2 F_{iz}^2- F_{0M}^2
\right)+\cO(\lambda^{-1})
\right]
\nn\\
&
-
\displaystyle\frac{aN_c}{2}\int d^4 x dz \,
\left[\,\frac{\lambda}{2}
\, \widehat{F}_{MN}^2+
\left(
-\frac{z^2}{6} \widehat{F}_{ij}^2+z^2 \widehat{F}_{iz}^2
-\widehat{F}_{0M}^2
\right)+\cO(\lambda^{-1})\right]\ ,
\label{actionrescaled}
\end{eqnarray}
while the total equations of motion
are
\begin{eqnarray}
&& D_M F_{0M}+\frac{1}{64\pi^2 a}
\epsilon_{MNPQ}\widehat{F}_{MN}F_{PQ}
+\cO(\lambda^{-1})=0 \ ,
\label{eq1}\\
& & D_N F_{MN}+\cO(\lambda^{-1})=0 \ .
\label{eq2}
\\
&& \del_M\widehat{F}_{0M}+\frac{1}{64\pi^2 a}
\epsilon_{MNPQ}\left\{
\tr(F_{MN}F_{PQ})+\half\widehat{F}_{MN}\widehat{F}_{PQ}
\right\}+\cO(\lambda^{-1})=0 \ ,
\label{eq3}\\
&& \del_{N}\widehat{F}_{MN}+\cO(\lambda^{-1})=0 \ .
\label{eq4}
\end{eqnarray}
Therefore, at the leading order, 
the warp factors $h(z)$ and $k(z)$
are approximated by $1$, so 
the $SU(2)$ part of the equations is nothing but the
standard YM equation in flat space. It is solved by a BPST instanton
located around $z\sim 0$. The electric 
$U(1)$ part is sourced by the instanton
density, as seen in (\ref{eq3}).
The explicit solution is \cite{HSSY}
\begin{align}
A_M^{\rm cl}=&-if(\xi)g\del_M g^{-1} \ ,~~
\wh A_0^{\rm cl}=
\frac{1}{8\pi^2 a}
\frac{1}{\xi^2}
\left[1-\frac{\rho^4}{(\rho^2+\xi^2)^2}
\right]\ ,~~~
A_0=\wh A_M=0 \ ,
\label{HSSYsol}
\end{align}
with the BPST instanton profile
\begin{align}
f(\xi)=\frac{\xi^2}{\xi^2+\rho^2}\ ,~~~
g(x)=\frac{(z-Z)+i(\vec{x}-\vec{X})\cdot\vec\tau}{\xi} \ ,~~
\xi=
\sqrt{(z-Z)^2+|\vec{x}-\vec{X}|^2} \ .
\label{defg}
\end{align}
$\rho$ is the size of the instanton, while 
$X^M=(X^1,X^2,X^3,Z)=(\vec X,Z)$ is  the position
of the soliton in the four-dimensional space.

Quantization of this soliton has been carried out in
Ref.~\citen{HSSY}. It is basically the same as the quantization of a
YM instanton in the moduli space approximation \cite{GeSa,Manton}, 
except for the additional potential in the moduli space
induced by the presence of the subleading terms in the action
(\ref{actionrescaled}). 
The moduli space for a single YM instanton 
is $\cM_1\simeq \bR^4\times \bR^4/\bZ_2$ parameterized by $(\vec X,Z)$
and $y^I$ ($I=1,2,3,4$) with the $\bZ_2$ action
$y^I\ra -y^I$. The radial component $\rho\equiv\sqrt{(y^I)^{2}}$ 
of $y^I$ gives the instanton size and the angular components
 $a^I=y^I/\rho$ parameterize
 the $SU(2)$ orientation of the instanton.
The quantization of the soliton is described by quantum mechanics
on this moduli space, with the Lagrangian 
\begin{eqnarray}
 L=\frac{m_X}{2}\dot{\vec{X}}^2+\frac{m_Z}{2}\dot Z^2+
\frac{m_y}{2} (\dot{y}^I)^2-U(\rho,Z)\ ,
\label{L1}
\end{eqnarray}
where the ``mass'' for each moduli is given by
\begin{eqnarray}
m_X=m_Z=\frac{m_y}{2}=8\pi^2a N_c\ ,  
\end{eqnarray}
and the potential
\begin{eqnarray}
 U(\rho,Z)=M_0 + 8\pi^2 a N_c
\left(
\frac{\rho^2}{6}
+\frac{1}{5(8\pi^2 a)^2}\frac{1}{\rho^2}
+\frac{Z^2}{3}
\right) 
\label{urhoz}
\end{eqnarray}
is obtained by substituting the solution (\ref{HSSYsol}) to the action
(\ref{actionrescaled}). $M_0\equiv 8\pi^2 \kappa$ is the classical mass at the leading
order in the $1/\lambda$ expansion.
The potential is classically minimized at 
\begin{eqnarray}
\rho_{\rm cl}^2
=\frac{1}{8\pi^2 a}\sqrt{\frac{6}{5}}\ ,
~~~
Z_{\rm cl}=0\ ,
\label{rhocl}
\end{eqnarray}
which shows that in fact the soliton has the
size of order $\lambda^{-1/2}$
when it is rescaled back to the original coordinates by \eqref{rescalemod}.
The Hamiltonian is given by
\begin{eqnarray}
 H=
\frac{-1}{2m_X}\left(\frac{\del}{\del \vec{X}}\right)^2
+
\frac{-1}{2m_Z}\left(\frac{\del}{\del Z}\right)^2
+\frac{-1}{2m_y}\left(\frac{\del}{\del y^I}\right)^2
+U(\rho,Z)\ .
\label{Ham}
\end{eqnarray}
This system has an $SO(4)\simeq (SU(2)_I\times SU(2)_J)/\bZ_2$
rotational symmetry acting on $y^I$.
Here $SU(2)_I$ and $SU(2)_J$ are interpreted as the isospin and spin
rotations, respectively, and they act on
 $\by\equiv y^4+iy^a\tau^a$ as
\begin{eqnarray}
 \by\ra g_I\by g_J
\end{eqnarray}
with $(g_I,g_J)\in SU(2)_I\times SU(2)_J$. The isospin and spin
operators are given by
\begin{eqnarray}
&&I^a=\frac{i}{2}\left(
y^4\frac{\del}{\del y^a}-y^a\frac{\del}{\del y^4}
-\epsilon_{abc}\,y^b\frac{\del}{\del y^c}
\right)\ ,\nn\\
&&J^a=\frac{i}{2}\left(
-y^4\frac{\del}{\del y^a}+y^a\frac{\del}{\del y^4}
-\epsilon_{abc}\,y^b\frac{\del}{\del y^c}
\right)\ ,
\label{IJ}
\end{eqnarray}
respectively. From this, we have $\vec I^2=\vec J^2$ and, hence, only
baryons with $I=J$ appear in this approach.

Quantum states of the baryon can be labeled using quantum numbers of
isospin/spin $I=J\equiv l/2 \ ,\,(l=1,3,5,\cdots)$, 
the eigenvalues of the third components of
isospin and spin operators $I^3$ and $J^3$,
and the quantum numbers $n_\rho=0,1,2,\cdots$ and $n_z=0,1,2,\cdots$,
which label the excitation numbers
of (almost) harmonic oscillators in $\rho$  and $Z$, respectively.
For example, the proton and neutron
have quantum numbers 
$(l,I_3,n_\rho,n_z)=(1,1/2,0,0)$ and $(l,I_3,n_\rho,n_z)=(1,-1/2,0,0)$, respectively.
The corresponding wavefunctions are
normalized spin/isospin states \cite{ANW}
\begin{eqnarray}
&&  |p\uparrow\rangle = \frac{1}{\pi} (y^1+iy^2)/\rho \ , \quad
  |p\downarrow\rangle = -\frac{i}{\pi} (y^4-iy^3)/\rho \ ,  
\nonumber\\
&&  |n\uparrow\rangle = \frac{i}{\pi} (y^4+iy^3)/\rho\ , \quad
  |n\downarrow\rangle = -\frac{1}{\pi} (y^1-iy^2)/\rho \ ,
\label{wf}
\end{eqnarray}
multiplied by the following $\rho$ and $Z$ wavefunctions, 
\begin{eqnarray}
 R(\rho)=\rho^{\tilde{l}}
e^{-\frac{m_y \omega_\rho}{2}\rho^2}\ ,~~
\psi_Z(Z)=e^{-\frac{m_Z \omega_Z}{2}Z^2}\ ,
\label{rzwave}
\end{eqnarray}
with $\tilde{l}=-1+2\sqrt{1+N_c^2/5}$, 
$\omega_\rho = 1/\sqrt{6}$, and $\omega_Z = \sqrt{2/3}$. 
The functions 
$R(\rho)$ and $\psi_Z(Z)$ should be multiplied by normalization factors.

\subsection{Our strategy}
\label{sec:2-2}

Our strategy for the calculation of the nuclear force consists of 
three steps:
\begin{itemize}
\item[1)] Construction of generic two-baryon solution of the YMCS
	 theory (\ref{model}),
\item[2)] Computation of the quantum-mechanical 
Hamiltonian for the moduli parameters, and 
\item[3)] Evaluation of the Hamiltonian with specified nucleon states.
\end{itemize}
This is a direct generalization of the single-baryon case to the
two-baryon case. 
In the following, we describe each step in more detail.

\subsubsection{Construction of two-baryon solution}
\label{sec:2-2-1}

The case of two baryons, which is our concern, can be considered by
solving the equations of motion of the original action (\ref{model})
with the constraint that the instanton number is 2. 
As we have seen, the rescaled variables are useful for seeing
the properties of the single baryon. There, one can start
with a BPST instanton solution in flat space, since the size of the
instanton is smaller than the scale of the curved background
geometry.
When we have two baryons, the situation is different. If the
distance between the two is larger than ${\cal O}(1/M_{\rm KK})$\footnote{
As a reference, if we use the value of $M_{\rm KK}$ that is fixed
by the rho meson mass, we have $1/M_{\rm KK}\simeq 0.208~{\rm fm}$. }
(or $\cO(\sqrt\lambda/\Mkk)$ in the rescaled coordinate), the
effect of the curved space-time comes into play, thus a similar analysis
cannot be performed. In this paper, we concentrate on the case where the
two baryons are close to each other, {\it i.e.},~the distance is smaller
than ${\cal O}(1/M_{\rm KK})$.

It is well-known that one can explicitly construct generic 
two-instanton solutions of Euclidean four-dimensional 
YM theory in flat space. We use ADHM construction
of the instantons for our purpose. The construction is reviewed in
Appendix~\ref{appB}. 

The two-instanton moduli space is parameterized by four quaternionic
parameters $(\bX_1,\bX_2,$ $\by_1,\by_2)$.
We summarize our notation for the quaternion in Appendix~\ref{appA}.
The quaternion has a representation by $2\times 2$ complex matrices
as in \eqref{cpxrep}. In this notation, these moduli parameters
can be written as
\begin{eqnarray}
\bX_i=Z_i+i\vec X_i\cdot\vec\tau\ ,~~
\by_i=y^4_i+i\vec y_i\cdot\vec\tau\ ,~~(i=1,2)
\end{eqnarray}
with $\vec X_i=(X_i^1,X_i^2,X_i^3)$
and $\vec y_i=(y_i^1,y_i^2,y_i^3)$. When the separation between
the two instantons is large, the two-instanton solution
can be approximated with
a superposition of two one-instanton configurations with
moduli parameters $(\bX_i,\by_i)$ ($i=1,2$).
Here, $X^M_i=(\vec X_i,Z_i)$ corresponds to the position of
 the instanton in the four-dimensional space,
 $\rho_i\equiv\sqrt{y_i^I y_i^I}$ is the size, and
 $\ba_i\equiv \by_i/\rho_i$ is the $SU(2)$ orientation of the instanton.

Defining $r^M\equiv X_1^M-X_2^M$ ($M=1,2,3,z$) as 
the relative position of the two
instantons and $|\br|=\sqrt{r^Mr^M}$ as the distance between them in
the four-dimensional space, the requirement for the flat space
approximation to be valid amounts to 
\begin{eqnarray}
|\br|<{\cal O}\left(
\sqrt{\lambda}/M_{\rm KK}
\right) \ .
\label{rg}
\end{eqnarray}
Note that this is written in the rescaled coordinates
(\ref{rescalemod}). 

As seen from the structure of the equations of motion in the $1/\lambda$
expansion, the only nonzero quantities at leading order are $A_M$ and 
$\widehat{A}_0$, as in the case of the single baryon. 
The equation of motion (\ref{eq3}) shows that the $U(1)$ part of the gauge
field is again sourced by the instanton density, now with two maxima at
the location of the separated baryons.
The explicit solution of the $SU(2)$ two-instanton solution and the $U(1)$
part will be presented in \S\ref{sec:3}, 
with the help of the ADHM construction
reviewed in Appendix~\ref{appB}.

\subsubsection{Calculation of quantum-mechanical Hamiltonian for 
two-baryon moduli}
\label{sec:2-2-2}

The next task is to obtain the classical potential 
$U(y_1^I, y_2^I, \vec{X}_1-\vec{X}_2, Z_1, Z_2)$. We substitute 
the two-instanton configuration into the action (\ref{actionrescaled}).
As we mentioned, the nonzero fields at the leading order are only 
$A_M(x)$ (the spatial components of the $SU(2)$) and 
$\widehat{A}_0(x)$ (the temporal component of the $U(1)$). Therefore, 
in the rescaled action (\ref{actionrescaled}), nonzero contributions
are 
\begin{eqnarray}
&&U= 2 M_0 +H_{\rm pot}^{(SU(2))} + H_{\rm pot}^{(U(1))}
+ {\cal O}\left(\lambda^{-1}\right)\ , 
\label{pottotal}
\\
&&
H_{\rm pot}^{(SU(2))}\equiv 
aN_c\displaystyle\int d^3 x dz \,\tr\left[\,
-\frac{z^2}{6} F_{ij}^2+z^2 F_{iz}^2
\right] 
=\frac{aN_c}{6}\displaystyle\int d^3 x dz \,\tr\left[\,
z^2F_{MN}^2
\right] 
\ , 
\label{su2pot}
\\
&&
H_{\rm pot}^{(U(1))}\equiv 
\displaystyle\frac{aN_c}{2}\int d^3 x dz \,
\left[\,\widehat{F}_{0M}^2
\right] \ .
\label{u1pot}
\end{eqnarray}
The first term of the total potential (\ref{pottotal}) is
the energy contribution from the first term in the action
\eqref{actionrescaled}, which gives
the leading order term in the $1/\lambda$ expansion.
The important part is the subleading order potential, 
$H_{\rm pot}^{(SU(2))}$  and $H_{\rm pot}^{(U(1))}$, which, in the case of
the single baryon, was used to determine the size of the instanton
classically and
was responsible for the quantum effect. We compute these for
the cases with two baryons, with
explicit dependence on the moduli parameters.
In the last equation of (\ref{su2pot}), we have used the self-dual
equation for the instanton.

The computation of $H_{\rm pot}^{(SU(2))}$ is straightforward although it
is technically involved, which will be presented in \S\ref{sec:3-1}. 
On the other hand, it turns out that $H_{\rm pot}^{(U(1))}$ is not easy to
compute. One can evaluate it numerically, but numerical results are not
useful for our purpose, as we will 
later need to make a moduli integration of it with the
baryon wavefunction.
Thus, we need the explicit analytic expression for the 
moduli dependence of the potential. For this purpose, 
we concentrate on the case with 
a large inter-baryon distance, 
\begin{eqnarray}
 {\cal O}(1/M_{\rm KK}) < |\br|\ ,
\label{rl}
\end{eqnarray}
where the left-hand side is the size of the single instanton, 
(\ref{rhocl}), in the rescaled coordinate. 
In this region, since there is only a slight overlap of
the instantons, one can obtain an analytic expression for 
$H_{\rm pot}^{(U(1))}$. The evaluation will be presented in
\S\ref{sec:3-2}. 

Together with the constraint (\ref{rg}), in this paper, we consider the
separation of the baryon satisfying 
\begin{eqnarray}
 {\cal O}
\left(1/M_{\rm KK}\right)< |\br| 
<  {\cal O}(\sqrt{\lambda}/M_{\rm KK})
\ 
\label{regionr}
\end{eqnarray}
in the rescaled coordinates (\ref{rescalemod}).

The quantum mechanics of the moduli parameters consists of the potential
term $U$ and the kinetic term. 
The kinetic term of the quantum mechanics of the moduli is given by the
moduli space metric. As opposed to the single-instanton case, the moduli
space metric of the two-instanton configuration is
complicated. 
It is found  that the asymptotic form in the case of a large
separation takes the form
\begin{eqnarray}
 ds^2 = ds_0^2 + ds_1^2 + {\cal O}(|\br|^{-3})\ ,
\end{eqnarray}
where $ds_0^2 = 2(dy^I_1)^2 + (dX^M_1)^2 + 2(dy^I_2)^2 + (dX^M_2)^2$
is just two copies of the metric for the single instanton, and 
$ds_1^2$ is ${\cal O}(|\br|^{-2})$,
which is our concern. This contributes to
the quantum mechanics as a ${\cal O}(|\br|^{-2})$ 
correction to the kinetic
term of the moduli dynamics. We explicitly compute this correction
in \S\ref{sec:3-3}. 

All together, the analytic expressions of the ${\cal O}(|\br|^{-2})$ 
terms
of the quantum mechanics are computed, and this is the 
Hamiltonian for the two-baryon interaction. The total expression is
summarized in \S\ref{sec:3-4}. 

\subsubsection{Evaluation of nucleon-nucleon potential}
\label{sec:2-2-3}

The final step is to evaluate the energy with the Hamiltonian of the
quantum mechanics. The interaction Hamiltonian is given in $1/|\br|$
expansion, and we treat this as a perturbation. 
The state of our baryons is specified at infinite separation, 
and we use simply the
tensor product of two copies of a single-baryon wavefunction.
This can be justified at leading order in the perturbation of quantum
mechanics.\footnote{
For describing the deuteron system, this perturbation is not the way to
proceed. One needs a minimum of the whole potential of moduli 
including $\br$, to obtain quantized energy of a bound
state of two baryons. Our interest in this paper is the potential force 
appearing in the scattering process of the two baryons. 
}

Since we have an analytic expression for the Hamiltonian, the
integration of the moduli parameter with the given wavefunctions is 
straightforward. The result is the nucleon-nucleon potential,
in particular if we choose the baryon wavefunctions to be that of a
nucleon. This integration will be presented in \S\ref{sec:4}.
The distance between the baryons in the three-dimensional space,
$|\vec{r}|=\!\!\sqrt{(X_1^1\!-\!\!X_2^1)^2\! +\! 
(X_1^2\!-\!\!X_2^2)^2\! +\! (X_1^3\!-\!\!X_2^3)^2}$, 
is related to the four-dimensional distance $|\br|$
as $|\br|=\sqrt{|\vec{r}|^2 + (Z_1-Z_2)^2}$. 
Note that we fix one of the moduli $|\vec{r}|$ 
and perform the integration of
the other moduli, $Z_1, Z_2, y_1^I, y_2^I$. This is because we are
interested in the potential in the scattering problem, rather than
the computation of the bound state energy.

Our final result for the nucleon-nucleon potential is given in 
(\ref{fresult0}) and (\ref{fresult1}). 
The central force (\ref{fresult0}) shows 
that the nucleons have a repulsive core. 
We also obtain the tensor force (\ref{fresult1}).
All the potentials have the form $|\vec{r}|^{-2}$, which is peculiar to
four-dimensional space, as described in the introduction.

To illustrate the properties of our nuclear force (\ref{fresult0}) 
and (\ref{fresult1}),
we next compute the
one-boson-exchange potential among nucleons. In our previous paper
\cite{HSS}, we derived the nucleon-nucleon-meson coupling in the D4-D8
model of the holographic QCD. By using this coupling, the summing up of
all types of mesons propagating among the nucleons should provide 
a certain aspect of the nuclear force. This computation can be carried out for 
arbitrary distances between the nucleons, as long as the nucleon radii do
not overlap each other. The resultant nuclear force, at larger
distances, exhibits the standard properties of the nuclear force, such
as scalar/tensor forces due to pion/$\rho$-meson/other-meson exchanges.
The computation will be presented in \S\ref{sec:5}.

We will find there that, in the region (\ref{regionr}), 
this one-boson-exchange potential does not
coincide with our nuclear force (\ref{fresult0}) and (\ref{fresult1})
derived using the ADHM construction of
two instantons. The reason is that when nucleons are close to each other
the nucleon itself is deformed by the effect of the other nucleon. In
deriving the one-meson-exchange potential, this effect cannot be taken
into account. Thus, naive computation based on the one-boson exchange is
not sufficient to capture the complete picture of the nuclear force at short
distances. 

\section{Effective Hamiltonian for two baryons}
\label{sec:3}

In this section, we calculate the effective Hamiltonian for the quantum
mechanics of the two-baryon state following the strategy described
in the previous section. The system is described as a quantum mechanics
of a particle living in the two-instanton moduli space.

\subsection{Potential from $SU(2)$ part}
\label{sec:3-1}

Let us first evaluate the contribution of the $SU(2)$ part of the
gauge field to the potential.
As explained in \S\ref{sec:2}, the leading term in the
$1/\lambda$ expansion can be obtained by substituting the two-instanton
solution in the flat space-time into the action \eqref{actionrescaled}.
The two-instanton solution can be obtained using the ADHM construction.
See Appendices~\ref{appA} and \ref{appB} for our notation
and a brief review of the ADHM construction.

The leading term in the $SU(2)$ part is obtained by evaluating \eqref{su2pot}.
This integral can be calculated using the formula \eqref{result1}
obtained in Appendix~\ref{app:SU2}. Substituting \eqref{n1k2} into
\eqref{result1}, we obtain
\begin{eqnarray}
 \int d^3 x dz \, z^2 \tr F_{MN}^2=8\pi^2
\left(\rho_1^2+\rho_2^2+2(Z_1^2+Z_2^2)+4(w^4)^2 \right)\ ,
\label{potSU2}
\end{eqnarray}
where
$w^4$
is the real part of \eqref{w}, which is given by
\begin{align}
w^4&=
-\frac{\vec r}{|\br|^2}
\cdot(\vec y_1y_2^4 -\vec y_2y_1^4 +\vec y_2\times \vec y_1)
=
\frac{\rho_1\rho_2}{2}\frac{r^a}{|\br|^2}
\tr\left(i\tau^a \ba_2^{-1}\ba_1\right)\ .
\end{align}

Therefore the potential from the $SU(2)$ part is given by
\begin{align}
H_{\rm pot}^{(SU(2))}&=
\frac{4\pi^2 aN_c}{3}
\left(
\rho_1^2+\rho_2^2+2(Z_1^2+Z_2^2)+4(w^4)^2
\right)\nn\\
&=\frac{4\pi^2 aN_c}{3}
\left(
\rho_1^2+\rho_2^2+2(Z_1^2+Z_2^2)+\rho_1^2\rho_2^2
\frac{r^ar^b}{|\br|^4}
\tr(i\tau^a\ba_2^{-1}\ba_1)
\tr(i\tau^b\ba_2^{-1}\ba_1)
\right)\ .
\label{HpotSU2}
\end{align}
The leading $r$-independent terms reproduce the
contribution of $SU(2)$ part in the one-instanton potential
\eqref{urhoz} for the two instantons.
The next-to-leading term gives the interaction between
the two baryons. It is of order $1/|\br|^2$ as expected
in a five-dimensional gauge theory.

\subsection{Potential from $U(1)$ part}
\label{sec:3-2}

The field strength of the $SU(2)$ part of the gauge field
satisfies \eqref{trF2} and then it is easy to see that
\begin{eqnarray}
 \widehat{A}_0 = \frac{1}{32 \pi^2 a}  \square \log \det L
\label{A0sol}
\end{eqnarray}
is the regular solution of the
equation of motion \eqref{eq3}, which vanishes at infinity.
Here, $L$ is given by \eqref{constSp} and \eqref{n1k2}.
By using the constraint \eqref{yyrw}, it can be written as
\begin{align}
L(x)=\mat{f_1(x), e(x),e(x),f_2(x)
}\ ,
\label{L2}
\end{align}
where
\begin{align}
f_i(x)&=\rho_i^2+|\bx-\bX_i|^2+|\bw|^2\ ,~~(i=1,2)\\
e(x)&=(\by_1\cdot\by_2) + \left(\bw\cdot(\bX_1+\bX_2-2\bx)\right)\ .
\end{align}

We evaluate the energy contribution of this $U(1)$ gauge field
when the separation between the two instantons is large.
Then, it can be confirmed that the energy density is mainly
concentrated around $\bx\sim\bX_i$ ($i=1,2$), where
the two instantons are located.
When $\bx$ is close to $\bX_1$, we can make the expansion
\begin{eqnarray}
\bx\sim \bX_1, \quad |\bx-\bX_1|\ll |\bx-\bX_2|\ .
\end{eqnarray}
Without losing generality, we can choose $\bX_1=0$
and $\bX_2$ to be very far away from the origin, and so we choose $\bx$
around the origin. The order estimate is
\begin{eqnarray}
 |\bX_2| \sim  |\bx-\bX_2| \sim |\br|\equiv |\bX_2-\bX_1| \gg
 |\bx-\bX_1|\sim |\bx| \ .
\end{eqnarray}
We can evaluate the matrix $L$ in this expansion, and obtain the
following expression: 
\begin{align}
\square \log\det L &=
\frac{4}{|\bx|^2} \left(1-\frac{\rho_1^4}{(|\bx|^2+\rho_1^2)^2}\right)
\nonumber \\
&
~~+ \frac{4}{|\br|^2}
\left(1 + \frac{2(\by_1\cdot \by_2)^2 \rho_1^2}{(|\bx|^2+\rho_1^2)^3}\right)
-\frac{8\rho_1^2}{(|\bx|^2+\rho_1^2)^3 }|\bw|^2
+ {\cal O} (|\br|^{-3}) \ .
\end{align}
In this expression, note that $\bw = {\cal O}(|\br|^{-1})$.
More explicitly, from \eqref{w}, we have 
\begin{eqnarray}
 |\bw|^2 = \frac{1}{|\br|^2}|\by_2\times\by_1|^2\ .
\end{eqnarray}
Then, the gauge field around the position of one of the two instantons
$\bx\sim \bX_1$ is expanded as
\begin{eqnarray}
\widehat{A}_0=\frac{1}{8\pi^2 a}
\left[\frac{1}{|\bx|^2} \left(1-\frac{\rho_1^4}{(|\bx|^2+\rho_1^2)^2}\right)
+ 
\frac{1}{|\br|^2}
\left(1 + \frac{2 Y \rho_1^2}{(|\bx|^2+\rho_1^2)^3}\right)
+ {\cal O} (|\br|^{-3})\right]\ ,
\nn\\
\label{expandAzero}
\end{eqnarray}
where
\begin{eqnarray}
Y\equiv (\by_1\cdot \by_2)^2-|\by_2\times\by_1|^2
=2 (\by_1\cdot \by_2)^2-|\by_1|^2|\by_2|^2
=\rho_1^2\rho_2^2\left(
2 (\ba_1\cdot \ba_2)^2-1
\right)\ .
\nn\\
\end{eqnarray}
The leading term in the $1/|\br|$ expansion reproduces
the result \eqref{HSSYsol} for the single instanton.
We are interested in the next-to-leading term of order $|\br|^{-2}$.
Note that the first term in the next-to-leading terms, 
\begin{eqnarray}
 \frac{1}{|\br|^2}\cdot 1\ ,
\end{eqnarray}
is precisely the leading contribution
of the five-dimensional Coulomb interaction with the
second instanton located at $\bx\sim \bX_2$. 

Let us compute the potential energy, with this expression for the gauge
field. The $U(1)$ part of the energy \eqref{u1pot} can be written as
\begin{eqnarray}
H_{\rm pot}^{(U(1))}
= - \frac{aN_c}{2} \int d^3x dz \widehat{A}_0 \square
\widehat{A}_0 \ .
\label{ABA}
\end{eqnarray}
Note that to obtain this expression we performed a partial integration,
it is harmless at this stage because the gauge field $\widehat{A}_0$ 
decays fast asymptotically. Then, we substitute the expanded expression
\eqref{expandAzero} to this energy formula.
This procedure is slightly ambiguous, since the integration
does not commute with the $1/|\br|$ expansion in \eqref{expandAzero}.
We give a more systematic way of evaluating the integral in
Appendix~\ref{app:U1}. Here, we present an easy way to obtain the correct
answer.
Substituting the expansion \eqref{expandAzero} into \eqref{ABA},
we obtain
\begin{eqnarray}
 \frac{N_c}{40\pi^2 a}
\left(\frac{1}{\rho_1^2}+\frac{1}{\rho_2^2}
\right)
\end{eqnarray}
as the leading term. This is just the sum of the energy contribution
of the single instanton in \eqref{urhoz}.
The next-to-leading order that
we are interested in is the cross terms, 
\begin{eqnarray}
&&
- \frac{aN_c}{2}\int d^3x dz \frac{1}{(8\pi^2 a)^2}
\left[
\left(
\frac{1}{|\bx|^2} \left(1-\frac{\rho_1^4}{(|\bx|^2+\rho_1^2)^2}\right)
\right)
\square
\left(
\frac{1}{|\br|^2}
\left(1 + \frac{2Y
 \rho_1^2}{(|\bx|^2+\rho_1^2)^3}\right)
\right)
\right]
\nonumber \\
&&
- \frac{aN_c}{2}\int d^3x dz \frac{1}{(8\pi^2 a)^2}
\left[
\left(
\frac{1}{|\br|^2}
\left(1 + \frac{2Y\rho_1^2}{(|\bx|^2+\rho_1^2)^3}\right)
\right)
\square
\left(
\frac{1}{|\bx|^2} \left(1-\frac{\rho_1^4}{(|\bx|^2+\rho_1^2)^2}\right)
\right)
\right]
\nonumber \\
&&
+ (\by_1 \leftrightarrow \by_2)\ .
\label{ABA2}
\end{eqnarray}
Here, $(\by_1\lra\by_2)$ denotes the contribution from the integration
around $\bx\sim\bX_2$, which is obtained by exchanging
$\by_1$ and $\by_2$ in the first and second terms of \eqref{ABA2}.
Note that the first term in \eqref{ABA2} is different from the second term.
This is because, at this stage, the partial integration suffers from a
surface term, owing to the constant $1/|\br|^2$.
Anyway, we can perform the integration analytically, and the result is
\begin{align}
H_{\rm pot}^{(U(1))}\simeq
 \frac{N_c}{40\pi^2 a}
\left(\frac{1}{\rho_1^2}+\frac{1}{\rho_2^2}
\right)
+
 \frac{N_c}{8\pi^2 a}
\frac{1}{|\br|^2}
\left[
\frac{1}{2} + \frac{2(\ba_1\cdot\ba_2)^2-1}{5} \left(\frac{\rho_2^2}{\rho_1^2}
+ \frac{\rho_1^2}{\rho_2^2}\right)
\right]
+\cO(|\br|^{-3})\ .
\label{HU1}
\end{align}

\subsection{Kinetic term}
\label{sec:3-3}

The kinetic term of the quantum mechanics of the two-baryon states
is given by
\begin{eqnarray}
\frac{m_X}{2} g_{\alpha\beta} \dot X^\alpha\dot X^\beta\ ,
\end{eqnarray}
where $X^\alpha=(X_i^M,y_i^I)$ ($\alpha=1,2,\dots,16$)
are the coordinates of the two-instanton moduli space
that are promoted to time-dependent variables and
$\dot X^\alpha=\frac{d}{dt}X^\alpha$. $g_{\alpha\beta}$ is
the metric of the two-instanton moduli space.

The line element of the two-instanton moduli space
for large separation is obtained in Ref.~\citen{PeZa} as
\begin{eqnarray}
ds^2=ds_0^2+ds_1^2+\cO(|\br|^{-3})
\end{eqnarray}
where
\begin{align}
ds_0^2&=
(d\bX_1\cdot d\bX_1)+(d\bX_2\cdot d\bX_2)+
2(d\by_1\cdot d\by_1)+2(d\by_2\cdot d\by_2)\ ,\nn\\
ds_1^2&=
\frac{2}{|\br|^2}\Big[
\rho_2^2(d\by_1\cdot d\by_1)+\rho_1^2(d\by_2\cdot d\by_2)
+2(\by_1\cdot d\by_1)(\by_2\cdot d\by_2)
\nn\\
&~~~-(\by_2\cdot d\by_1)^2-(\by_1\cdot d\by_2)^2
-2(\by_1\cdot \by_2)(d\by_1\cdot d\by_2)\nn\\
&~~~+2\,\epsilon_{IJKL}\,y_1^Iy_2^Jdy_1^{K}dy_2^{L}
-\left((\by_2\cdot d\by_1)-(\by_1\cdot d\by_2)\right)^2
\,
\Big] \ .
\label{metric}
\end{align}
The leading terms $ds_0^2$ in \eqref{metric} is
just a sum of the metric for each instanton, which
gives the canonical kinetic term \eqref{L1}.
The next-to-leading terms $ds_1^2$ contribute to the
potential of order $1/|\br|^2$.

The kinetic term of the Hamiltonian is given by
the Laplacian $\nabla^2$ of the two-instanton moduli space as
\begin{eqnarray}
H_{\rm kin}=-\frac{1}{2m_X}\nabla^2 \ .
\label{Hkin}
\end{eqnarray}
The outline of the calculation is summarized
in Appendix~\ref{app:metric}. The result is
\begin{eqnarray}
\nabla^2&=&\nabla_0^2+\nabla_1^2+\cO(|\br|^{-3})\ ,
\end{eqnarray}
where
\begin{align}
\nabla_0^2&=
\left(\frac{\del}{\del X_1^M}\right)^2
+\left(\frac{\del}{\del X_2^M}\right)^2
+\half\left(\frac{\del}{\del y_1^I}\right)^2
+\half\left(\frac{\del}{\del y_2^I}\right)^2\ ,\\
\nabla_1^2&=-\frac{1}{|\br|^2}\Bigg[
\frac{\rho_2^2}{2}
\left(\frac{\del}{\del y_1^I}\right)^2
+\frac{\rho_1^2}{2}
\left(\frac{\del}{\del y_2^I}\right)^2
-\left(y_2^I\frac{\del}{\del y_1^I}\right)^2
-\left(y_1^I\frac{\del}{\del y_2^I}\right)^2
+y_1^I\frac{\del}{\del y_1^I}
+y_2^I\frac{\del}{\del y_2^I}
\nn\\
&~~~+\,\epsilon_{IJKL}\,y_1^Iy_2^J
\frac{\del}{\del y_1^K}\frac{\del}{\del y_2^L}
+\left(y_1^I y_2^J+y_1^Jy_2^I-(\by_1\cdot\by_2)\,\delta^{IJ}
\right)\frac{\del}{\del y_1^I}\frac{\del}{\del y_2^J}
\,\Bigg]\ ,
\label{lap}
\end{align}
Using the spin/isospin operators \eqref{IJ},
it can also be written as
\begin{align}
\nabla_0^2
&=
\left(\frac{\del}{\del X_1^M}\right)^2
+\left(\frac{\del}{\del X_2^M}\right)^2
+\half\sum_{i=1,2}\left[
\frac{1}{\rho_i^3}
\frac{\del}{\del\rho_i}
\left(\rho_i^3\frac{\del}{\del\rho_i}\right)
-\frac{4}{\rho_i^2}I_i^aI_i^a
\right]\ ,\\
\nabla_1^2=&
-\frac{1}{|\br|^2}\Bigg[
\frac{\rho_2^2}{2}\left(
\frac{1}{\rho_1^3}
\frac{\del}{\del\rho_1}
\left(\rho_1^3\frac{\del}{\del\rho_1}\right)
-\frac{4}{\rho_1^2}I_1^aI_1^a
\right)
+
\frac{\rho_1^2}{2}\left(
\frac{1}{\rho_2^3}
\frac{\del}{\del\rho_2}
\left(\rho_2^3\frac{\del}{\del\rho_2}\right)
-\frac{4}{\rho_2^2}I_2^aI_2^a
\right)\nn\\
&
+4I_1^aI_2^a
+\rho_1\rho_2\frac{\del}{\del\rho_1}
\frac{\del}{\del \rho_2}
+\rho_1\frac{\del}{\del \rho_1}
+\rho_2\frac{\del}{\del \rho_2}
-\left(y_2^I \frac{\del}{\del y_1^I}\right)^2
-\left(y_1^I \frac{\del}{\del y_2^I}\right)^2
\,\Bigg]\ .
\label{lap2}
\end{align}
Here, $a,b,c=1,2,3$ are the indices for the
$SU(2)$ adjoint representation
and the subscripts $i=1,2$ label the two instantons.

\subsection{Summary}
\label{sec:3-4}

By collecting \eqref{HpotSU2}, \eqref{HU1}, and
\eqref{Hkin} with \eqref{lap} or \eqref{lap2},
the total Hamiltonian is obtained as
\begin{eqnarray}
 H =
 H_0 + H_1.
\label{Ham2}
\end{eqnarray}
The leading order Hamiltonian $H_0$ is just two copies of
the Hamiltonian for one baryon \eqref{Ham} obtained in
Ref.~\citen{HSSY}
\begin{align}
 H_0 =\sum_{i=1,2}\Bigg[
&
\frac{-1}{2m_X}
\left(\frac{\partial}{\partial\vec X_i}\right)^2
+\frac{-1}{2m_y}
\left(\frac{\partial}{\partial y_i^I}\right)^2
+\frac{-1}{2m_Z}
\left(\frac{\partial}{\partial Z_i}\right)^2
+U(\rho_i,Z_i)
\Bigg]\ ,
\label{hssypot}
\end{align}
where $U(\rho_i,Z_i)$ is the potential given in \eqref{urhoz}.

The final term $H_1$ gives the $\cO(|\br|^{-2})$
interaction between the two baryons,
\begin{eqnarray}
 H_1 = H_1^{(U(1))} + H_1^{(SU(2))} - \frac{1}{2m_X}\nabla^2_1 \ ,
\label{h1}
\end{eqnarray}
where $m_X=8\pi^2a N_c$ and 
\begin{eqnarray}
&& 
H_1^{(U(1))} = 
 \frac{N_c}{8\pi^2 a}
\frac{1}{|\br|^2}
\left[
\frac{1}{2} + \frac{2(\ba_1\cdot\ba_2)^2-1}{5} 
\left(\frac{\rho_2^2}{\rho_1^2}  
+\frac{\rho_1^2}{\rho_2^2}  \right)
\right] \ ,
\label{h1u1}
\\
&& 
H_1^{(SU(2))}=
\frac{4\pi^2 aN_c}{3}
\rho_1^2\rho_2^2
\frac{r^ar^b}{|\br|^4}
\tr(i\tau^a\ba_2^{-1}\ba_1)
\tr(i\tau^b\ba_2^{-1}\ba_1)
\ ,
\label{h1su2}
\end{eqnarray}
and $\nabla_1^2$ is defined in (\ref{lap2}).

\section{Nucleon-nucleon interaction}
\label{sec:4}

We are ready for the evaluation of the nucleon-nucleon interaction
potential at short distances, using the quantum mechanical Hamiltonian 
(\ref{Ham2})
obtained in the previous section. This section is devoted to
demonstrating these manipulations in detail. 

We evaluate the interaction Hamiltonian $H_1$ with the nucleon wave
function obtained in Ref.~\citen{HSSY}. The wavefunctions for consistent
quantum states of two asymptotic nucleons are given in
\S\ref{sec:4-1}, and with the wavefunctions, the computation of
the expectation value of $H_1$ follows in \S\ref{sec:4-2}. 
Finally, the decomposition of $\langle H_1 \rangle$ into a central force
and a tensor force is given in \S\ref{sec:4-3}. Our final result
for the nucleon-nucleon potential at short distances is
(\ref{fresult0}) and (\ref{fresult1}).

\subsection{Wavefunctions of two-nucleon states}
\label{sec:4-1}

We have two baryons, and the wavefunction for them 
is given by a tensor product of the wavefunctions (\ref{wf}) and
(\ref{rzwave}). 
We can arrange (\ref{wf}) in a simple form,
\begin{eqnarray}
\frac1{\pi} (\tau^2 \ba)_{IJ}
=
\left(
\begin{array}{cc}
|p\uparrow\rangle &  |p\downarrow\rangle \\
 |n\uparrow\rangle &  |n\downarrow\rangle
\end{array}
\right)_{IJ} \ , 
\label{aIJ}
\end{eqnarray}
where $\tau^2$ is the Pauli matrix $\tau^a$ with $a=2$.
Here, we specify the matrix element using the indices $I,J$ that take
values in $\{\pm 1/2\}$.
{}From \eqref{aIJ}, we see that
$(I,J)$ component of the matrix $\tau^2 \ba$ is directly related
to the wavefunction of the spin/isospin state with $(I^3,J^3)=(I,J)$.
For the two nucleons specified by $(I^3_1, J^3_1, I^3_2, J^3_2)$,
our wavefunction is
\begin{eqnarray}
 \frac{1}{\pi^2} (\tau^2 \ba_1)_{I_1J_1} (\tau^2 \ba_2)_{I_2 J_2} \ ,
\label{awave}
\end{eqnarray}
where we have omitted the upper index $3$ in $I_i^3$ and $J_i^3$
for simplicity.

The wavefunction for the instanton size variable $\rho$ is given by
a tensor product of (\ref{rzwave}), which is
$R(\rho_1)R(\rho_2)$.
To reduce the computational effort for the evaluation of the $\rho$
integral, 
we use the following simplified wavefunction instead of 
\eqref{rzwave},
\begin{eqnarray}
R(\rho_1)R(\rho_2) \ , \quad \mbox{where}\quad 
 R(\rho) \equiv 
 \rho^{\tilde{l}}\exp\left[-\rho^2
\right] \ .
\label{rrhofun}
\end{eqnarray}
Hereafter, we mean $R(\rho)$ using this expression.
This is obtained by a rescaling 
$(1/2)m_y \omega_\rho \rho^2 \to \rho^2$.
Note that, among the terms in $H_1$ (\ref{h1}), 
$H_1^{(U(1))}$ and $\nabla_1^2$ are
invariant under the rescaling, so we can just replace the wavefunction
by this simplified one. However, $H_1^{(SU(2))}$ has a scaling dimension
that needs to be taken into account.
The overall normalization of (\ref{rrhofun}) will be taken care of later.

The wavefunction for the $Z$ modulus, for the nucleon states 
of $n_Z=0$, is again a tensor product of (\ref{rzwave})
 and given by
\begin{eqnarray}
\psi_Z(Z_1)\psi_Z(Z_2)\ .
\label{wavefunz}
\end{eqnarray}
This is, again, up to a normalization constant.

\subsection{Spin/isospin matrix elements of interaction Hamiltonian}
\label{sec:4-2}

Let us evaluate 
the expectation value of the interaction Hamiltonian $H_1$,
for given wavefunctions for bra and ket, (\ref{awave}), (\ref{rrhofun}),
and (\ref{wavefunz}). Since the moduli $Z_1$ and $Z_2$ are decoupled
from the rest, we treat them later independently. 
Here, first, we integrate the moduli $y^I_1$ and $y^I_2$.
The integration measure is 
$d^4y_1 d^4y_2 = \rho_1^3 \rho_2^3 d\rho_1 d\rho_2 d\Omega_1 d\Omega_2$,
where $d\Omega_i$ is the integration over the angular coordinates
$a_i^I$ in $SU(2)\sim S^3$. 
The normalization is given by $\int d\Omega_i = 2\pi^2$, which is the
volume unit $S^3$. For the integral $d\Omega_i$, 
the following formulas for the integration over $g\in SU(2)$ are useful: 
\begin{eqnarray}
&& 
\int \! d\Omega \; g_{ij} g^{-1}_{kl} g_{mn} g^{-1}_{pq}
= \frac{\pi^2}{3} 
\left[
2\left(
\delta_{il} \delta_{mq} \delta_{jk}\delta_{np}
+\delta_{iq} \delta_{ml} \delta_{jp}\delta_{nk}
\right)
-\delta_{il} \delta_{mq} \delta_{jp}\delta_{nk}
-\delta_{jk}\delta_{np}\delta_{iq} \delta_{ml}
\right] \ ,
\nonumber \\
&& 
\int \! d\Omega \; g_{ij} g^{-1}_{kl} 
= \pi^2 \delta_{il} \delta_{jk} \ .
\label{formulag}
\end{eqnarray}


\vspace{3mm}

\noindent
\underline{Evaluation of $H_1^{(U(1))}$}

\vspace{3mm}

Let us evaluate the expectation value of $H_1^{(U(1))}$ 
(\ref{h1u1}). 
In $H_1^{(U(1))}$, the spatial coordinates $X^M_i$ do not couple 
to $y^I_i$, so it provides only a central force, and does not yield 
a tensor force. 

As for the integration over the $\rho$ variables, 
only the following
three integrals appear:
\begin{eqnarray}
 \int_0^\infty \! d\rho \; \rho^3 R(\rho)^2 \ , \quad
 \int_0^\infty \! d\rho \; \rho^5 R(\rho)^2 \ , \quad
 \int_0^\infty \! d\rho \; \rho R(\rho)^2 \ .
\end{eqnarray}
On the other hand, to compute the expectation values, we need to include
the normalization of the wavefunctions. Considering both, we only need
the following ratios for our computation: 
\begin{eqnarray}
\left(
 \int_0^\infty \! d\rho \; \rho^5 R(\rho)^2
\right)/
\left(
  \int_0^\infty \! d\rho \; \rho^3 R(\rho)^2 
\right)
=1+\frac{\tilde{l}}{2} \ ,
\nonumber \\
\left(
 \int_0^\infty \! d\rho \; \rho R(\rho)^2
\right)/
\left(
  \int_0^\infty \! d\rho \; \rho^3 R(\rho)^2
\right)
=\frac{2}{1+\tilde{l}} \ .
\label{ratios}
\end{eqnarray}
These can be derived by partial integrations. Using these, we obtain
\begin{eqnarray}
\left\langle
\displaystyle\frac{\rho_2^2}{\rho_1^2}+\frac{\rho_1^2}{\rho_2^2}
\right\rangle
=
\frac{
 \displaystyle\int d\rho_1 d\rho_2 \; \rho_1^3 \rho_2^3
\left(
\displaystyle\frac{\rho_2^2}{\rho_1^2}+\frac{\rho_1^2}{\rho_2^2}
\right) R(\rho_1)^2R(\rho_2)^2}
{
 \int d\rho_1 d\rho_2 \; \rho_1^3 \rho_2^3
R(\rho_1)^2R(\rho_2)^2}
=2 \; \frac{\tilde{l}+2}{\tilde{l}+1} \ .
\label{rhores1}
\end{eqnarray}
Note that we could use (\ref{rrhofun}) because 
$\frac{\rho_2^2}{\rho_1^2}+\frac{\rho_1^2}{\rho_2^2}$ is
scale-invariant. 

Next, let us perform the $\ba$ integration. In $H_1^{(U(1))}$, 
the nontrivial term is only $(\ba_1\cdot\ba_2)^2$, so we compute the matrix
element of this. We first note
\begin{eqnarray}
 (\ba_1\cdot\ba_2)^2 = \frac14 
\;{\rm tr}\!\left[
\ba_1^\dagger \ba_2
\right]
{\rm tr}\!\left[
\ba_1\ba_2^\dagger
\right] 
= \frac14 (\ba_1^\dagger)_{KL} (\ba_1)_{MN} (\ba_2)_{LK} 
(\ba_2^\dagger)_{NM} \ .
\end{eqnarray}
Thus, with the wavefunction (\ref{awave}), the matrix element is
\begin{eqnarray}
&&
\left\langle 
 (\ba_1\cdot\ba_2)^2
\right\rangle_{(I'_1,J'_1, I'_2, J'_2),(I_1,J_1,I_2,J_2)}
\nonumber 
\\
&&
= \frac{1}{4\pi^4}\!\!
\int \!\! d\Omega_1 \!\! \int \!\! d\Omega_2
\;
 (\ba_1^\dagger \tau^2)_{J'_1I'_1} (\ba_2^\dagger \tau^2)_{J'_2I'_2} 
(\ba_1^\dagger)_{KL} (\ba_1)_{MN} 
(\ba_2)_{LK}  (\ba_2^\dagger)_{NM} 
 (\tau^2 \ba_1)_{I_1J_1} (\tau^2 \ba_2)_{I_2J_2} \ . \quad
\nn\\
\label{aaint}
\end{eqnarray}
Here, we have used 
$((\tau^2 \ba)_{I' J'})^* = (\ba^\dagger \tau^2)_{J'I'}$.
For the integral (\ref{aaint}), we can use the formula (\ref{formulag})
to obtain
\begin{align}
&
\left\langle 
 (\ba_1\cdot\ba_2)^2
\right\rangle_{(I'_1,J'_1, I'_2, J'_2),(I_1,J_1,I_2,J_2)}
\nn\\
&
= \frac{1}{18}
\left[
5\delta_{I'_1I_1}\delta_{I'_2 I_2} \delta_{J'_1 J_1} \delta_{J'_2 J_2} 
- \delta_{I'_1I_2}\delta_{I'_2 I_1} \delta_{J'_1 J_1} \delta_{J'_2 J_2} 
- \delta_{I'_1I_1}\delta_{I'_2 I_2} \delta_{J'_1 J_2} \delta_{J'_2 J_1} 
+2 \delta_{I'_1I_2}\delta_{I'_2 I_1} \delta_{J'_1 J_2} \delta_{J'_2 J_1} 
\right] \nn\\
&
= \frac{1}{4}
\delta_{I'_1I_1}\delta_{I'_2 I_2} \delta_{J'_1 J_1} \delta_{J'_2 J_2} 
+\frac{1}{36}
(\vec\tau_{I_1'I_1}\cdot\vec\tau_{I_2'I_2})
(\vec\tau_{J_1'J_1}\cdot\vec\tau_{J_2'J_2})\ .
\label{a1a2}
\end{align}
To obtain the last expression of \eqref{a1a2}, we have used the
identity
\begin{eqnarray}
2\delta_{I'_1I_2}\delta_{I'_2 I_1} 
= 
\delta_{I'_1I_1}\delta_{I'_2 I_2} 
+ \vec\tau_{I'_1I_1}\cdot\vec\tau_{I'_2 I_2} \ .
\label{f1}
\end{eqnarray}

Combining \eqref{a1a2} with (\ref{rhores1}), we obtain the matrix element for
$H_1^{(U(1))}$ as
\begin{align}
&
\left\langle 
H_1^{(U(1))}
\right\rangle_{(I'_1,J'_1, I'_2, J'_2),(I_1,J_1,I_2,J_2)}
\nonumber
\\
&=\frac{N_c}{8\pi^2 a}\frac{1}{|\br|^2}
\left[\frac{1}{10}\frac{3\tilde l+1}{\tilde l+1}
\delta_{I'_1I_1}\delta_{I'_2 I_2} \delta_{J'_1 J_1} \delta_{J'_2 J_2} 
+\frac{1}{45}\frac{\tilde{l}+2}{\tilde{l}+1}
(\vec\tau_{I_1'I_1}\cdot\vec\tau_{I_2'I_2})
(\vec\tau_{J_1'J_1}\cdot\vec\tau_{J_2'J_2})
\right]\ .
\label{A}
\end{align}
The $Z_1,Z_2$ dependence in $|\br|^2$ will be integrated later,
together with the other terms in $H_1$.


\vspace{3mm}

\noindent
\underline{Evaluation of $H_1^{(SU(2))}$}

\vspace{3mm}

Next, we consider (\ref{h1su2}). 
First we make the integration over $\rho_1$ and $\rho_2$. 
In the present case, the integral is not invariant under the scaling of
$\rho$ as opposed to the previous examples. This time, after the scaling
$ (1/2)m_y \omega_\rho \rho^2 \rightarrow \rho^2$,
we obtain the multiplicative new factor $4/(m_y \omega_\rho)^2$.
Therefore, using (\ref{ratios}), we obtain
\begin{align}
\left\langle 
\rho_1^2 \rho_2^2
\right\rangle
=
\frac{4}{(m_y \omega_\rho)^2}
\frac{\int d\rho_1 \; \rho_1^5 R(\rho_1)^2}{\int d\rho_1 \; \rho_1^3 R(\rho_1)^2}
 \frac{\int d\rho_2 \; \rho_2^5 R(\rho_2)^2}{\int d\rho_2 \; \rho_2^3 R(\rho_2)^2}
=
\frac{4}{(m_y \omega_\rho)^2}
\left(
1+ \frac{\tilde{l}}{2}
\right)^2 \ .
\label{rint2}
\end{align}

Next, we consider the integration over $\ba_1$ and $\ba_2$.
It is obvious that it  
proceeds in the same manner as $H_1^{(U(1))}$. 
To use the formulas (\ref{formulag}), 
we bring the relevant part of $H_1^{(SU(2))}$ to the form
\begin{eqnarray}
\tr(i\tau^a\ba_2^{-1}\ba_1)
\tr(i\tau^b\ba_2^{-1}\ba_1)
&=&
\tr(\tau^a\ba_2^{-1}\ba_1)
\tr(\tau^b\ba_1^{-1}\ba_2) 
\nonumber 
\\
&=&
(\ba_1)_{ij} (\ba_1^{-1})_{pq} 
(\ba_2)_{qr} (\ba_2^{-1})_{ki} (\tau^a)_{jk} (\tau^b)_{rp} \ .
\end{eqnarray}
Then, using \eqref{formulag}, we obtain
\begin{align}
&
\left\langle 
\tr(i\tau^a\ba_2^{-1}\ba_1)
\tr(i\tau^b\ba_2^{-1}\ba_1)
\right\rangle_{(I'_1,J'_1, I'_2, J'_2),(I_1,J_1,I_2,J_2)}
\nn\\
&=\, \frac{1}{9}
\left[
2\delta_{ab}(
4\delta_{I'_1I_1}\delta_{I'_2 I_2} \delta_{J'_1 J_1} \delta_{J'_2 J_2} 
\!+\!
\delta_{I'_1I_2}\delta_{I'_2 I_1} \delta_{J'_1 J_1} \delta_{J'_2 J_2} 
)
\right.
\nonumber \\
& \hspace{20mm}
\!+\! (\tau^a\tau^b)_{J'_1 J_1}
(\delta_{I'_1I_1}\delta_{I'_2 I_2} \delta_{J'_2 J_2} 
\!-\!2 \delta_{I'_1I_2}\delta_{I'_2 I_1} \delta_{J'_2 J_2} 
)
\nonumber \\
& \hspace{30mm}
~~+ (\tau^b\tau^a)_{J'_2 J_2}
(\delta_{I'_1I_1}\delta_{I'_2 I_2} \delta_{J'_1 J_1} 
-2 \delta_{I'_1I_2}\delta_{I'_2 I_1} \delta_{J'_1 J_1} 
)
\nonumber \\
& \hspace{40mm}
\left.
+ (\tau^a)_{J'_1 J_2} (\tau^b)_{J'_2 J_1}
(
4\delta_{I'_1I_2}\delta_{I'_2 I_1} 
-2\delta_{I'_1I_1}\delta_{I'_2 I_2} 
)
\right] \ .
\label{aaaa-1}
\end{align}
To simplify this expression, we use \eqref{f1} and
the Fierz transformation
\begin{align}
& 2(\tau^a)_{J'_1 J_2} (\tau^b)_{J'_2 J_1}
- (\tau^a\tau^b)_{J'_1 J_1}\delta_{J'_2 J_2} 
- (\tau^b\tau^a)_{J'_2 J_2}\delta_{J'_1 J_1} 
\nonumber
\\
&
\quad 
= (\tau^a)_{J'_2 J_2} (\tau^b)_{J'_1 J_1}
+(\tau^b)_{J'_2 J_2} (\tau^a)_{J'_1 J_1}
-\delta_{ab}(\tau^c)_{J'_2 J_2} (\tau^c)_{J'_1 J_1}
-\delta_{ab}\delta_{J'_1 J_1} \delta_{J'_2 J_2}  \ .
\label{f2}
\end{align}
Then, \eqref{aaaa-1} becomes
\begin{align}
&
\left\langle 
\tr(i\tau^a\ba_2^{-1}\ba_1)
\tr(i\tau^b\ba_2^{-1}\ba_1)
\right\rangle_{(I'_1,J'_1, I'_2, J'_2),(I_1,J_1,I_2,J_2)}
\nn\\
&=\,
\delta_{ab}
\delta_{I'_1I_1}\delta_{I'_2 I_2} \delta_{J'_1 J_1} \delta_{J'_2 J_2} 
\nn\\
&
+
\frac{1}{9}
(\vec\tau_{I'_1I_1}\cdot\vec\tau_{I'_2 I_2})
\left(
(\tau^a)_{J'_2 J_2} (\tau^b)_{J'_1 J_1}
\!+\!(\tau^b)_{J'_2 J_2} (\tau^a)_{J'_1 J_1}
\!-\!\delta_{ab}(\vec\tau_{J'_2 J_2}\cdot\vec\tau_{J'_1 J_1})
\right)
 \ .
\label{aint2}
\end{align}
Combining (\ref{rint2}) and (\ref{aint2}), we arrive at
\begin{align}
&
\left\langle 
H_1^{(SU(2))}
\right\rangle_{(I'_1,J'_1, I'_2, J'_2),(I_1,J_1,I_2,J_2)}
\nn\\
&=\, \frac{(\tilde{l}+2)^2}{32 \pi^2 a N_c} \frac{|\vec r|^2}{|\br|^4}
\biggm[
\delta_{I'_1I_1}\delta_{I'_2 I_2} \delta_{J'_1 J_1} \delta_{J'_2 J_2} 
\nn\\
& \hspace{10mm}
+
\frac{1}{9}
(\vec\tau_{I'_1I_1}\cdot\vec\tau_{I'_2 I_2} )
\left(2
(\hat{\vec r}\cdot\vec\tau)_{J'_2 J_2} 
(\hat{\vec r}\cdot\vec\tau)_{J'_1 J_1}
-\vec\tau_{J'_2 J_2}\cdot\vec\tau_{J'_1 J_1}
\right)
\biggm]
 \ ,
\label{omegaterm}
\end{align}
where $\hat{\vec r} \equiv \vec{r}/|\vec{r}|$.


\vspace{3mm}

\noindent
\underline{Evaluation of $\nabla_1^2$}

\vspace{3mm}

Finally, we evaluate $\nabla_1^2$ in $H_1$. 
In (\ref{lap2}), only the last two terms
$\left(y_1^I \frac{\partial}{\partial y_2^I}\right)^2 
+ \left(y_2^I \frac{\partial}{\partial y_1^I}\right)^2$ 
have the angular dependence. Thus, as for the terms other than these,
we only have to perform the $\rho$ integral.
For example, using 
\begin{eqnarray}
 \frac{1}{\rho^3}\frac{\partial}{\partial\rho}
\left(\rho^3 \frac{\partial}{\partial\rho}R(\rho)\right)
=
\left(4\rho^2 + (-8-4\tilde{l}) + (\tilde{l}^2 
+ 2\tilde{l})\frac{1}{\rho^2}
\right)
R(\rho) \ ,
\end{eqnarray}
the ratio formulas (\ref{ratios}) can be applied, and 
this results in
\begin{eqnarray}
&&\left\langle 
\nabla_1^2 
\right\rangle_{(I'_1,J'_1, I'_2, J'_2),(I_1,J_1,I_2,J_2)}
\nonumber \\
&&\hspace{10mm}
= 
\frac{1}{|\br|^2}
\left(
\frac{(\tilde{l}+2)(\tilde{l}+5)}{\tilde{l}+1}
\delta_{I'_1I_1}\delta_{I'_2 I_2} 
-\vec\tau_{I'_1I_1}\cdot\vec\tau_{I'_2 I_2}
\right) \delta_{J'_1 J_1} \delta_{J'_2 J_2} 
\nonumber \\
&&\hspace{20mm}
+
\frac{1}{|\br|^2}
\left\langle
\left[
\left(y_1^I \frac{\partial}{\partial y_2^I}\right)^2
+ \left(y_2^I \frac{\partial}{\partial y_1^I}\right)^2 
\right] 
\right\rangle_{(I'_1,J'_1, I'_2, J'_2),(I_1,J_1,I_2,J_2)} \ .
\label{nablaeva}
\end{eqnarray}
We have used 
$4I_1^a I_2^a = \vec\tau_{I'_1I_1} \cdot\vec\tau_{I'_2 I_2}$.

To perform the integral in the last term in (\ref{nablaeva}),
we first perform a partial integration
\begin{eqnarray}
&& \int \! d^4y_1 \int \! d^4y_2 \; 
(\psi')^*_{(I'_1,J'_1, I'_2, J'_2)}
\left(y_1^I \frac{\partial}{\partial y_2^I}\right)^2
\psi_{(I_1,J_1,I_2,J_2)}
\nonumber 
\\
&&~~
= - 
 \int \! d^4y_1 \int \! d^4y_2 \; 
\left(
y_1^I \frac{\partial}{\partial y_2^I}
\psi'_{(I'_1,J'_1, I'_2, J'_2)}
\right)^*
y_1^I \frac{\partial}{\partial y_2^I}
\psi_{(I_1,J_1,I_2,J_2)} \ .
\label{partiali}
\end{eqnarray}
Using the wavefunctions (\ref{awave}) and (\ref{rrhofun}), we find
\begin{align}
 y_2^I \frac{\partial}{\partial y_1^I}
\psi_{(I_1,J_1,I_2,J_2)} 
 =\, &
 \frac{1}{\pi^2} 
\left[
(\tau^2 \ba_2)_{I_1J_1} (\tau^2 \ba_2)_{I_2 J_2}
\frac{\rho_2}{\rho_1}R(\rho_1)R(\rho_2)
\right.
\nonumber \\
&
\left.
\quad +
(\tau^2 \ba_1)_{I_1J_1} (\tau^2 \ba_2)_{I_2 J_2}
(\ba_1\cdot \ba_2) \; \rho_1 \rho_2
\frac{\partial}{\partial \rho_1}\left(\frac{R(\rho_1)}{\rho_1}\right)
R(\rho_2)
\right] \ .
\end{align}
Thus, in (\ref{partiali}), the following integrals for $\rho$ appear:
\begin{eqnarray}
&& 
\frac{\int \rho_1^3 d\rho_1 \int \rho_2^3 d\rho_2 \;
\frac{\rho_2^2}{\rho_1^2} R(\rho_1)^2 R(\rho_2)^2}
{\int \rho_1^3 d\rho_1 \int \rho_2^3 d\rho_2 \;
R(\rho_1)^2 R(\rho_2)^2}
= \frac{\tilde{l}+2}{\tilde{l}+1} \ ,
\\
&&
\frac{\int \rho_1^3 d\rho_1 \int \rho_2^3 d\rho_2 \;
\rho_2^2 R(\rho_2)^2 R(\rho_1) 
\frac{\partial}{\partial \rho_1}\left(\frac{R(\rho_1)}{\rho_1}\right)}
{\int \rho_1^3 d\rho_1 \int \rho_2^3 d\rho_2 \;
R(\rho_1)^2 R(\rho_2)^2}
= (-2)\frac{\tilde{l}+2}{\tilde{l}+1} \ ,
\\
&&
\frac{\int \rho_1^3 d\rho_1 \int \rho_2^3 d\rho_2 \;
\rho_1^2\rho_2^2 R(\rho_2)^2 
\left[
\frac{\partial}{\partial \rho_1}\left(\frac{R(\rho_1)}{\rho_1}\right)
\right]^2}
{\int \rho_1^3 d\rho_1 \int \rho_2^3 d\rho_2 \;
R(\rho_1)^2 R(\rho_2)^2}
= \frac{(\tilde{l}+2)(\tilde{l}+5)}{\tilde{l}+1} \ .
\end{eqnarray}
Here, again, we have used (\ref{ratios}). Then, straightforward
calculations with the formulas \eqref{formulag} and \eqref{a1a2}  show that
\begin{align}
&
 \left\langle
\left[
\left(y_1^I \frac{\partial}{\partial y_2^I}\right)^2
+ \left(y_2^I \frac{\partial}{\partial y_1^I}\right)^2 
\right] 
\right\rangle_{(I'_1,J'_1, I'_2, J'_2),(I_1,J_1,I_2,J_2)} 
\nonumber 
\\
&~~
= 
-\frac{(\tilde{l}+2)(\tilde{l}+5)}{\tilde{l}+1} 
\left[
\half\delta_{I'_1I_1}\delta_{I'_2 I_2} \delta_{J'_1 J_1} \delta_{J'_2 J_2} 
+\frac{1}{18}
(\vec\tau_{I'_1I_1}\cdot\vec\tau_{I'_2 I_2})
(\vec\tau_{J'_1J_1}\cdot\vec\tau_{J'_2 J_2})
\right] \ .
\label{Aprime}
\end{align}

As a result, the matrix elements of the third term in \eqref{h1} are
obtained from \eqref{nablaeva} and \eqref{Aprime} as
\begin{align}
&-\frac{1}{2m_X}\vev{\nabla_1^2\,}_{(I'_1,J'_1, I'_2, J'_2),(I_1,J_1,I_2,J_2)} 
\nn\\
&~~=\,\frac{1}{32\pi^2 aN_c|\br|^2}\Bigg[
~2(\vec\tau_{I'_1I_1}\cdot\vec\tau_{I'_2 I_2})
\delta_{J'_1 J_1} \delta_{J'_2 J_2} 
\nn\\
&~~~~~~~~
-\frac{(\tilde l+2)(\tilde l+5)}{\tilde l+1}\left(
\delta_{I'_1I_1}\delta_{I'_2 I_2} \delta_{J'_1 J_1} \delta_{J'_2 J_2} 
-\frac{1}{9}
(\vec\tau_{I'_1I_1}\cdot\vec\tau_{I'_2 I_2})
(\vec\tau_{J'_1J_1}\cdot\vec\tau_{J'_2 J_2})
\right)
\Bigg]\ .
\label{nab1}
\end{align}
Since $\tilde l\sim\cO(N_c)$, this term is $\cO(1)$ in the $1/N_c$
expansion, while the other terms \eqref{A} and \eqref{omegaterm} are
$\cO(N_c)$.
Therefore, this term is subleading, compared with the contributions from
$H_1^{(U(1))}$ and $H_1^{(SU(2))}$.


\vspace{3mm}

\noindent
\underline{Summary of the result}

\vspace{3mm}

Summing up all the results (\ref{A}), (\ref{omegaterm}),
and (\ref{nab1}), we have the
spin/isospin matrix elements of the interaction Hamiltonian 
\begin{eqnarray}
 \left\langle
H_1
\right\rangle_{(I'_1,J'_1, I'_2, J'_2),(I_1,J_1,I_2,J_2)} 
=  c_1 \; \frac{1}{|\br|^2}
+ c_2 \; \frac{|\vec r|^2}{|\br|^4} \ , 
\label{AB}
\end{eqnarray}
where the coefficients are given by 
\begin{align}
c_1=\,&\frac{N_c}{8\pi^2 a}
\left[\frac{1}{10}\frac{3\tilde l+1}{\tilde l+1}
\delta_{I'_1I_1}\delta_{I'_2 I_2} \delta_{J'_1 J_1} \delta_{J'_2 J_2} 
+\frac{1}{45}\frac{\tilde{l}+2}{\tilde{l}+1}
(\vec\tau_{I_1'I_1}\cdot\vec\tau_{I_2'I_2})
(\vec\tau_{J_1'J_1}\cdot\vec\tau_{J_2'J_2})
\right]
\nn\\
&+\frac{1}{32\pi^2 aN_c}\Bigg[
~2(\vec\tau_{I'_1I_1}\cdot\vec\tau_{I'_2 I_2})
\delta_{J'_1 J_1} \delta_{J'_2 J_2} 
\nn\\
&~~~~~~~
-\frac{(\tilde l+2)(\tilde l+5)}{\tilde l+1}\left(
\delta_{I'_1I_1}\delta_{I'_2 I_2} \delta_{J'_1 J_1} \delta_{J'_2 J_2} 
-\frac{1}{9}
(\vec\tau_{I'_1I_1}\cdot\vec\tau_{I'_2 I_2})
(\vec\tau_{J'_1J_1}\cdot\vec\tau_{J'_2 J_2})
\right)
\Bigg]\ ,
\label{c1}
\\
c_2=\,&
\frac{(\tilde{l}+2)^2}{32 \pi^2 a N_c}
\left[
\delta_{I'_1I_1}\delta_{I'_2 I_2} \delta_{J'_1 J_1} \delta_{J'_2 J_2} 
+
\frac{1}{9}
(\vec\tau_{I'_1I_1}\cdot\vec\tau_{I'_2 I_2} )
\left(2
(\hat{\vec r}\cdot\vec\tau)_{J'_2 J_2} 
(\hat{\vec r}\cdot\vec\tau)_{J'_1 J_1}
-\vec\tau_{J'_2 J_2}\cdot\vec\tau_{J'_1 J_1}
\right)
\right]
 \ .
\label{c2}
\end{align}

\subsection{Final result for nuclear force at short distance}
\label{sec:4-3}

We finally evaluate this Hamiltonian (\ref{AB})
with the wavefunction for $Z_i$, (\ref{wavefunz}), and take the leading
term in the large $N_c$ expansion. 

In our result (\ref{AB}), 
the $Z$ dependence is included in the four-dimensional
distance, $|\br|=\sqrt{|\vec{r}|^2 + (Z_1-Z_2)^2}$.
We fix the baryon distance $|\vec{r}|$ in the real space, 
and perform the
integration over $Z_1$ and $Z_2$.
To perform the integration, we rewrite the wavefunction
(\ref{wavefunz}) as
\begin{eqnarray}
\exp\left[-
\frac12 m_Z \omega_Z \left((Z_1)^2 \!+\! (Z_2)^2\right)
\right]
= \exp\left[-
 m_Z \omega_Z \left(((Z_1\! -\!Z_2)/2)^2 + ((Z_1\! +\!Z_2)/2)^2\right)
\right] \ .
\nn\\
\end{eqnarray}
Since (\ref{AB}) consists of two terms, $1/|\br|^2$ and $1/|\br|^4$,
we use the following formula for integration,
\begin{eqnarray}
&& \int_{-\infty}^\infty dx \frac{e^{-\beta^2 x^2}}{x^2 + \alpha^2}
= \frac{\pi e^{\alpha^2 \beta^2} (1-{\rm Err}[\alpha \beta])}{\alpha}
\ , \\
&& \int_{-\infty}^\infty dx \frac{e^{-\beta^2 x^2}}{(x^2 + \alpha^2)^2}
= \frac{2\sqrt{\pi} \alpha\beta +\pi (1-2\alpha^2\beta^2)
e^{\alpha^2 \beta^2} (1-{\rm Err}[\alpha \beta])}{2\alpha^3} \ .
\end{eqnarray}
Here Err is the error function,
\begin{eqnarray}
 {\rm Err}[x]\equiv \frac{2}{\sqrt{\pi}}\int_0^x\! e^{-t^2}\; dt \ .
\end{eqnarray}
Then we obtain
\begin{eqnarray}
&& \left\langle\frac{1}{|\br|^2}\right\rangle_Z
=
\frac{ \sqrt{\pi} \beta e^{\beta^2 |\vec{r}|^2/4} 
(1-{\rm Err}[\beta |\vec{r}|/2])}{2|\vec{r}|} 
\ , 
\\
&& \left\langle\frac{1}{|\br|^4}\right\rangle_Z
=
\beta
\frac{\beta |\vec{r}| +\sqrt{\pi} (1-\beta^2 |\vec{r}|^2/2)
e^{\beta^2 |\vec{r}|^2/4} (1-{\rm Err}[\beta |\vec{r}|/2])}{4|\vec{r}|^3}
\ ,
\end{eqnarray}
where 
$\beta = \sqrt{2 m_Z \omega_Z} = (16 (2/3)^{1/2}\pi^2 a N_c)^{1/2}$. 

For large $N_c$, the argument of the error function, $\beta |\vec{r}|$ is
very large for nonzero $|\vec{r}|$. We can use the following asymptotic
formula for the error function,
\begin{eqnarray}
 1-{\rm Err}(x) = \frac{1}{\sqrt{\pi}}e^{-x^2}\left[
\frac{1}{x}- \frac{1}{2x^3} + {\cal O}(x^{-4})
\right] \ .
\end{eqnarray}  
By using this expression, the $Z$-integral results are markedly
simplified,
\begin{eqnarray}
\left\langle\frac{1}{|\br|^2}\right\rangle_Z
= \frac{1}{|\vec r|^2}  + {\cal O}(1/N_c) \ , \quad
\left\langle\frac{1}{|\br|^4}\right\rangle_Z
= \frac{1}{|\vec r|^4} + {\cal O}(1/N_c) \ . \quad
\end{eqnarray}
This expression can be easily guessed since this is just
a substitution of $Z_1=Z_2=0$. The value $Z_1=Z_2=0$ is
the classical value of the location of the instanton in the $z$ space,
and the large $N_c$ limit should reproduce the classical result.

On the other hand, the coefficients $c_1$ and $c_2$ in (\ref{AB}) are
evaluated in the large $N_c$ expansion as
\begin{align}
& c_1 = \pi N_c \left(
\frac{81}{10}
\delta_{I'_1I_1}\delta_{I'_2 I_2} \delta_{J'_1 J_1} \delta_{J'_2 J_2} 
+ \frac{3}{5}
(\vec\tau_{I'_1 I_1}\cdot\vec\tau_{I'_2 I_2})
(\vec\tau_{J'_1 J_1}\cdot\vec\tau_{J'_2 J_2})
\right) + {\cal O}(1) \ ,
\\
&
c_2 = \pi N_c 
\frac35
\Bigm(
9
\delta_{I'_1I_1}\delta_{I'_2 I_2} \delta_{J'_1 J_1} \delta_{J'_2 J_2} 
-
(\vec\tau_{I'_1 I_1}\cdot\vec\tau_{I'_2 I_2})
(\vec\tau_{J'_1 J_1}\cdot\vec\tau_{J'_2 J_2})
\nonumber 
\\
&
\hspace{40mm}
\left.
+2
(\vec\tau_{I'_1I_1}\cdot\vec\tau_{I'_2 I_2} )
(\hat{\vec r}\cdot\vec\tau)_{J'_1 J_1}
(\hat{\vec r}\cdot\vec\tau)_{J'_2 J_2}
\right)
+ {\cal O}(1) \ .
\end{align}
Here, for deriving these, we 
used $\tilde{l}^2 = (4/5)N_c^2 + {\cal O}(1)$.
It is interesting that in fact all the contributions from $\nabla_1^2$
drop off, since they are subleading compared with the terms from 
$H_1^{(U(1))}$ and $H_1^{(SU(2))}$ in this large $N_c$ expansion.

Therefore, the leading term of the interaction Hamiltonian is
\begin{align}
&\left\langle
H_1
\right\rangle_{(I'_1,J'_1, I'_2, J'_2),(I_1,J_1,I_2,J_2)} 
\nn\\
&= \frac{\pi N_c}{\lambda|\vec r|^2}
\left(
\frac{27}{2}
\delta_{I'_1I_1}\delta_{I'_2 I_2} \delta_{J'_1 J_1} \delta_{J'_2 J_2} 
+\frac65
(\vec\tau_{I'_1I_1}\cdot\vec\tau_{I'_2 I_2})
(\hat{\vec r}\cdot\vec\tau)_{J'_2 J_2} 
(\hat{\vec r}\cdot\vec\tau)_{J'_1 J_1}
\right) \ .
\label{finalv}
\end{align}
We rescaled $\vec r$ back to the original coordinate (remember the
rescaling (\ref{rescalemod})). This is the nucleon-nucleon potential 
at a short distance in the large $N_c$ limit. 

Let us decompose this force into a central force $V_{\rm C}(|\vec{r}|)$
and a tensor force $V_{\rm T}(|\vec{r}|)$,
\begin{eqnarray}
&& V = V_{\rm C}(|\vec{r}|) + S_{12}V_{\rm T}(|\vec{r}|) \ .
\label{forcedec}
\end{eqnarray}
Here, the tensor operator $S_{12}$ is defined by
\begin{eqnarray}
S_{12}\equiv
 3 (\vec\sigma_1\cdot\hat{\vec{r}}\,)
(\vec\sigma_2\cdot\hat{\vec{r}}\,) - \vec\sigma_1\cdot \vec\sigma_2 
= 12 (\vec J_1\cdot\hat{\vec{r}}\,)
(\vec J_2\cdot\hat{\vec{r}}\,) - 4\vec J_1\cdot \vec J_2 \ ,
\label{S12}
\end{eqnarray}
 where 
$\vec\sigma_i=(\sigma_i^1,\sigma_i^2,\sigma_i^3)= 2\vec J_i$ are
the Pauli-spin operators
(it is just twice the spin
operator). 
Applying the decomposition (\ref{forcedec}) to our result
(\ref{finalv}), we obtain
\begin{eqnarray}
&& V_{\rm C}(|\vec{r}|) 
= 
\pi \left(
\frac{27}{2} + \frac{32}{5}
(I_1^a I_2^a)(J_1^bJ_2^b)
\right)
\frac{N_c}{\lambda} \frac{1}{|\vec{r}|^2}
\ , 
\label{fresult0}
\\
&& V_{\rm T}(|\vec{r}|) = 
\frac{8\pi}{5}I_1^a I_2^a 
\frac{N_c}{\lambda} 
\frac{1}{|\vec{r}|^2} 
\ .
\label{fresult1}
\end{eqnarray}

We have derived the nuclear force at a short distance. As already
mentioned, the expression is valid in the region (\ref{regionr}). 
We have found that there is a strong repulsive core in the central
force (\ref{fresult0}). Our finding is quite important,
as an analytic computation of the repulsive core of the nuclear
force, based on strongly coupled QCD.

The tensor force (\ref{fresult1}) is found to be negative for $I=0$
(since $I_1^a I_2^a = -3/4$), 
which is
consistent with the experimental observation and also with lattice QCD
calculations \cite{Ishii:2006ec}.
It is intriguing that the nuclear force has $|\vec{r}|^{-2}$ dependence.
This $|\vec{r}|^{-2}$ is peculiar to the physics with one extra spatial 
dimension, and thus it is 
a direct consequence of the holographic approach to QCD.
It would be interesting to fit the lattice and experimental
data of the nuclear force with our result.

\section{Comparison with one-boson-exchange potential}
\label{sec:5}

In this section, for a comparison, we
compute the nucleon-nucleon potential using the one-boson-exchange
picture. This can be done by integrating the meson propagator with
the nucleon-nucleon-meson coupling obtained in our previous paper
\cite{HSS} in the D4-D8 model in the holographic QCD.\footnote{In
\S\ref{sec:5} only, we use the upper (or lower) 
index ``$(1)$'' and ``$(2)$'' to label the gauge fields and the currents
for the two instantons, 
and we do not use the rescaled coordinates (\ref{rescalemod}).
This is for making use of
the notation of Ref.~\citen{HSS}.  } 
We find some
discrepancy between the two pictures. The reason for this
discrepancy is basically the fact that in the one-boson-exchange
picture the deformation of the nucleon by the other nucleon has not been
taken into account, while our potential obtained in \S\ref{sec:4}
includes this effect via the ADHM construction in the previous section.
In sum, we find that the one-boson-exchange potential captures merely a
part of the nucleon-nucleon potential.

\subsection{Interaction potential}
\label{sec:5-1}

In Ref.~\citen{HSS}, the nucleon-nucleon-meson couplings were computed,
by extracting the asymptotic behavior of the one-baryon 
solution.\footnote{In Refs.~\citen{Hong-Rho-Yee-Yi,PaYi}, the
couplings were computed by a different approach, using baryon spinor fields in the
bulk curved space-time.}
The one-boson-exchange potential is obtained by just evaluating
the energy of a superposition of two asymptotic baryons, which are regarded
as point particles sourcing the meson fields propagating between them.
The distance $r$ should be larger than the instanton size 
$\rho$, of course, 
because we use the asymptotic form of the solution (which was 
given in \S 2.3 in Ref.~\citen{HSS}) 
as a baryon solution. 

Before getting into the details of the evaluation of the potential 
energy of our system, it is instructive to
consider a simple example of a two-electron system in classical 
electrodynamics with the action
\begin{eqnarray}
 S = \int d^4x\left(-\frac{1}{4}F_{\mu\nu}^2-j^\mu A_\mu \right)\ ,
\end{eqnarray}
where $j_\mu$ is a current of an external source.
Here, we work in the Lorenz gauge $\del_\mu A^\mu=0$ and consider a
static configuration. Then, the equation of motion 
$\Delta A^\mu=j^\mu$
is solved by
\begin{eqnarray}
 A^\mu(\vec x)=(\Delta^{-1}j^\mu)(\vec x)
\equiv\int d^3y\, \Delta^{-1}(\vec x,\vec y)\, j^\mu(\vec y) \ ,
\end{eqnarray}
where
$\Delta^{-1}(\vec x,\vec y)\equiv\frac{-1}{4\pi}\frac{1}{|\vec x-\vec y|}$
is the Green's function for the Laplacian $\Delta$.
The on-shell action is evaluated as
\begin{eqnarray}
 S =-\half \int d^4x\,A^\mu\Delta A_\mu=-\half \int d^4x\,j^\mu A_\mu=
-\half \int d^4x\,j^\mu \Delta^{-1}j_\mu\ .
\label{onshell}
\end{eqnarray}
Let us consider the current associated with
two electrons placed at $\vec X_i$ ($i=1,2$):
\begin{eqnarray}
&&j_\mu=j^{(1)}_\mu+j^{(2)}_\mu\ ,\nn\\
&&j^{(i)}_0(\vec x)=e\delta^3(\vec x-\vec X_i)\ ,~~
 j^{(i)}_1=j^{(i)}_2=j^{(i)}_3= 0\ .~~(i=1,2)
\label{delta}
\end{eqnarray}
Then, the solution of the equation of motion is given by
\begin{eqnarray}
&& A_0(x)=A_0^{(1)}(x) + A_0^{(2)}(x) \ , \quad A_1=A_2=A_3=0 \ , \nn\\
&& A_0^{(i)} = e\Delta^{-1} (\vec x,\vec X_i)\ .~~(i=1,2)
\label{eoma0}
\end{eqnarray}
Substituting this in the on-shell action \eqref{onshell} and picking up
the cross term, we obtain the potential
due to the interaction of the two sources as
\begin{eqnarray}
V = \int d^3x\, A_\mu^{(1)} \Delta A^{(2)\mu}
  = \int d^3x\, A_\mu^{(1)} j^{(2)\mu}
  = -e^2\Delta^{-1}(\vec X_2,\vec X_1)\ .
\label{Vpot}
\end{eqnarray}

Let us follow this line of argument for our action (\ref{model}) in the
curved space-time. It was shown in Ref.~\citen{HSS} 
that nonlinear terms in the
equations of motion can be neglected in the asymptotic region.
Therefore, we are allowed to consider the linearized equations of motion 
in the curved space-time:
\begin{align}
h(z)\del_\mu^2\cA_i+\del_z (k(z)\del_z\cA_i)=0\ ,~~~
\del_\mu^2\cA_z+\del_z (h(z)^{-1}\del_z (k(z)\cA_z))=0\ ,
\label{linear}
\end{align}
with the gauge condition
\begin{eqnarray}
 h(z) \partial^\mu \cA_\mu + \partial_z (k(z)\cA_z)=0 \ .
\label{gauge}
\end{eqnarray}
The potential energy analogous to \eqref{Vpot} is 
\begin{eqnarray}
 V = 2\kappa \int d^3x dz 
\tr\left[
-A_0^{(1)} j_0^{(2)}
+A_i^{(1)} j_i^{(2)}
+A_z^{(1)} j_z^{(2)}
\right] \ ,
\label{hamilint}
\end{eqnarray}
where the ``current'' $j_\alpha^{(2)}$ is defined as
\begin{eqnarray}
&& j_0^{(2)} = \left(
h(z) \partial_i \partial_i
 + \partial_z k(z)\partial_z
\right)A_0^{(2)} \ ,
\nn\\
&& j_j^{(2)} = \left(
h(z) \partial_i \partial_i
 + \partial_z k(z)\partial_z
\right)A_j^{(2)} \ ,
\nn\\
&& j_z^{(2)} = k(z)\left(
\partial_i \partial_i
 + \partial_z h(z)^{-1}\partial_z k(z)
\right)A_z^{(2)} \ .
\label{j2}
\end{eqnarray}
Here, $A^{(i)}_\alpha$ ($i=1,2$) are the asymptotic solutions for a
single baryon located at $x^M\sim X_i^M$ that satisfy the linearized equations of
motion \eqref{linear} and the gauge condition \eqref{gauge}. They are
explicitly obtained in Ref.~\citen{HSS}.
The ``current'' \eqref{j2} behaves as the pointlike source placed at
$X_2^M$, which corresponds to the delta function source $j_\mu^{(2)}$
 in \eqref{delta} in
the previous example. Of course, the asymptotic solutions
can only be trusted in the asymptotic region, since the nonlinear terms
in the equations of motion will become important near the position of
the instanton $X_i^M$. However, here, we simply assume that the baryons can be
treated as point particles to obtain the one-boson-exchange potential.

{}For simplicity,
we consider the static configuration.\footnote{
This amounts to throwing away momentum-dependent nuclear force such as
$L\cdot S$ force. }
By noting that 
our focus is on the leading order behavior of the large
$\lambda$ and large $N_c$ limit,  the leading contribution 
to the currents turns out to come only from
\begin{eqnarray}
 \widehat{A}_0, A_i, A_z\ ,
\end{eqnarray}
which are of order ${\cal O}(\lambda^{-1})$ while
the next-to-leading 
order is by the component $A_0$, which is 
of order ${\cal O}(\lambda^{-1}N_c^{-1})$. All the 
other components are  of order ${\cal O}(\lambda^{-2}N_c^{-1})$. 
Thus, let us consider only the leading order components.

The solutions and the ``current'' are
\begin{eqnarray}
 &&
\wh{A}_0^{(1)} = \frac{-1}{2a\lambda}
G(\vec{x},z;\vec{X}_1,Z_1) \ ,
\nn\\
&&
\wh{j}_0^{(2)}= \frac{-1}{2a\lambda} 
\delta^3(\vec{x}-\vec{X}_2) \delta(z-Z_2) \ ,
\\
&&
\nn\\
&&
A_i^{(1)b} = -2\pi^2 (\rho_1)^2
{\rm tr}
\left(
\tau^b \ba_1\tau^a
(\ba_1)^{-1}
\right)
\left(
\epsilon^{iaj}\frac{\partial}{\partial X_1^j}
-\delta^{ia}\frac{\partial}{\partial Z_1}
\right)
G(\vec{x},z;\vec{X}_1,Z_1) \ ,
\nonumber
\\
&&
j_i^{(2)b}
=
-2\pi^2 (\rho_2)^2
{\rm tr}
\left(
\tau^b \ba_2\tau^a
(\ba_2)^{-1}
\right)
\left(
\epsilon^{iaj}\frac{\partial}{\partial X_2^j}
-\delta^{ia}\frac{\partial}{\partial Z_2}
\right)\delta^3(\vec{x}-\vec{X}_2) \delta(z-Z_2) \ ,
\\
&&
\nn\\
&&
A_z^{(1)b} = -2\pi^2 (\rho_1)^2
{\rm tr}
\left(
\tau^b \ba_1\tau^a
(\ba_1)^{-1}
\right)
\frac{\partial}{\partial X_1^a}
H(\vec{x},z;\vec{X}_1,Z_1) \ ,
\nonumber
\\
&&
j_z^{(2)b}
=
-2\pi^2 (\rho_2)^2
{\rm tr}
\left(
\tau^b \ba_2\tau^a
(\ba_2)^{-1}
\right)
\frac{\partial}{\partial X_2^a}
\delta^3(\vec{x}-\vec{X}_2) \delta(z-Z_2) \ .
\end{eqnarray}
Note that as in Ref.~\citen{HSS}, these expressions are written
in terms of 
the original variables without the rescaling (\ref{rescalemod}),
keeping them in the order of $\cO(1)$.
Here, $G$ and $H$ are the Green's functions associated with
the linearized equations of motion \eqref{linear}. They are
obtained in Ref.~\citen{HSS} as
\begin{eqnarray}
 G = \kappa \sum_{n=1}^\infty \psi_n(Z_2) 
\psi_n(Z_1)  Y_n(|\vec{r}|)  \ ,
\quad
 H = \kappa \sum_{n=0}^\infty \phi_n(Z_2) 
\phi_n(Z_1)  Y_n(|\vec{r}|) \ , 
\label{green:GH}
\end{eqnarray}
where $\{\psi_n\}_{n=1,2,\cdots}$ is a complete set
of the eigenfunctions satisfying the eigenequation
\begin{eqnarray}
 -h(z)^{-1}\del_z(k(z)\del_z\psi_n)=\lambda_n\psi_n\ ,
\end{eqnarray}
with eigenvalue $\lambda_n$ and the normalization condition
\begin{eqnarray}
 \kappa\int dz\,h(z)\psi_n\psi_m=\delta_{nm}\ ,
\end{eqnarray}
and $\{\phi_n\}_{n=0,1,\cdots}$ is defined as
\begin{eqnarray}
 \phi_n(z) = \frac{1}{\sqrt{\lambda_n}} \partial_z\psi_n(z) \ ,
\quad
\phi_0(z) = \frac{1}{\sqrt{\kappa\pi}} \frac{1}{k(z)} \ .
\end{eqnarray}
The eigenfunction $\psi_n(z)$ is an even (odd) function
for odd (even) $n$. When the five-dimensional gauge field
is expanded by $\{\psi_n(z)\}$, the coefficient fields
are interpreted as the vector (for odd $n$) and axial-vector
(for even $n$) meson fields, and the eigenvalues $\lambda_n$ are proportional to
the mass squared of the corresponding mesons. The function $Y_n(|\vec r|)$ 
is the Yukawa potential associated with the vector/axial-vector mesons
of mass $\sqrt{\lambda_n}$ :
\begin{eqnarray}
 Y_n(|\vec{r}|)= -\frac{1}{4\pi} 
\frac{e^{-\sqrt{\lambda_n}|\vec{r}|}}{|\vec{r}|} \ .
\label{Yn}
\end{eqnarray}

We substitute these expressions to the interaction potential 
(\ref{hamilint}) to obtain
\begin{align}
 V = &\,\kappa
\bigg[
-\frac{1}{4a^2 \lambda^2}
G(\vec{X}_2, Z_2;\vec{X}_1, Z_1)
\nonumber\\
&~~~+ 4\pi^4 (\rho_1)^2(\rho_2)^2
{\rm tr}
\left(
\tau^b \ba_1\tau^a
(\ba_1)^{-1}
\right)
{\rm tr}
\left(
\tau^b \ba_2\tau^c
(\ba_2)^{-1}
\right)
\nonumber\\
&~~~~~~~
\times 
\left(
\epsilon^{iaj}\frac{\partial}{\partial X_1^j}
-\delta^{ia}\frac{\partial}{\partial Z_1}
\right)
\left(
\epsilon^{ick}\frac{\partial}{\partial X_2^k}
-\delta^{ic}\frac{\partial}{\partial Z_2}
\right)
G(\vec{X}_2, Z_2;\vec{X}_1, Z_1)
\nonumber\\
&~~~+4\pi^4 (\rho_1)^2(\rho_2)^2
{\rm tr}
\left(
\tau^b \ba_1\tau^a
(\ba_1)^{-1}
\right)
{\rm tr}
\left(
\tau^b \ba_2\tau^c
(\ba_2)^{-1}
\right)
\times 
\frac{\partial}{\partial X_1^a}
\frac{\partial}{\partial X_2^c}
H(\vec{X}_2, Z_2;\vec{X}_1, Z_1)
\bigg] \ .
\label{hamilf}
\end{align}
This is the inter-baryon potential energy that we want to evaluate as the
nuclear force, in the one-boson-exchange approximation. 

\subsection{Short distance behavior}
\label{sec:5-4}

When the distance between the solitons is smaller than 
$1/M_{\rm KK}$, the Green's functions $G$ and $H$ can be approximated by
their flat-space analogue as explained in our previous paper \cite{HSS}, 
\begin{eqnarray}
 G = H = 
\frac{-1}{4\pi^2} 
\frac{1}{ |\vec{X}_1-\vec{X}_2|^2 + (Z_1-Z_2)^2 } \ .
\end{eqnarray}
Substituting this expression to the inter-instanton 
potential energy (\ref{hamilf}), we can easily see that only the first
term in (\ref{hamilf}) remains, while 
the other terms cancel each other and vanish, 
\begin{eqnarray}
  V = 
\frac{\kappa}{16\pi^2 a^2 \lambda^2}
\frac{1}{ |\vec{X}_1-\vec{X}_2|^2 + (Z_1-Z_2)^2 } \ .
\label{pot4d}
\end{eqnarray}
This is a harmonic potential in four-dimensional space. 
Once the $z$-part of the wavefunction for the nucleon,
(\ref{wavefunz}),  is taken into account, the expectation value is
given just by substituting the classical value $Z_1=Z_2=0$ to
the above potential (\ref{pot4d}),
\begin{eqnarray}
  V = 
\frac{27\pi}{2} \frac{N_c}{\lambda}
\frac{1}{|\vec{r}|^{2}} \ .
\label{pot4d2}
\end{eqnarray}
Here, $|\vec{r}|$ is the distance between the nucleons.
We find that only the central force appears.

This expression (\ref{pot4d2}) is apparently different from
our previous results (\ref{fresult0}) and (\ref{fresult1}).
In fact, in (\ref{pot4d2}), there is no
isospin dependence, and no tensor force. 
The reason is that the one-boson-exchange
description is not sufficient to capture the entire nuclear force. In fact,
the two-instanton solution in ADHM construction is not just a
superposition of two BPST instantons. In particular, there is a
deformation of the instanton that causes additional contribution. 
The one-boson-exchange potential (\ref{pot4d2}) only captures
a part of the complete results (\ref{fresult0}) and (\ref{fresult1})
obtained in the previous subsection.

In Appendix~\ref{sec:height}, we try to evaluate the potential height
for nucleons overlapping each other in real space, in the
one-boson-exchange picture.

\subsection{Large distance behavior}
\label{sec:5-2}

Let us look at the large distance behavior of the nucleon-nucleon
potential obtained from (\ref{hamilf}). The large distance means
$|\vec X_1- \vec X_2| \gg 1$ in the unit $M_{\rm KK}=1$. In this limit,
essentially only the pion dominates, since only the pion 
is the zero mode while others have Yukawa potential that decays
exponentially fast.

The contribution of the pion comes from $n=0$ component of the Green's
function $H$ defined in \eqref{green:GH}.
Therefore, in the potential (\ref{hamilf}), we are interested in the
last term, and the function $H$ is now approximated by a massless
Green's function in three dimensions:
\begin{align}
 V \simeq~ &
\kappa \cdot 4\pi^4 (\rho_1)^2(\rho_2)^2
{\rm tr}
\left(
\tau^b \ba_1\tau^a
(\ba_1)^{-1}
\right)
{\rm tr}
\left(
\tau^b \ba_2\tau^c
(\ba_2)^{-1}
\right)
\nonumber\\
&
\times 
\frac{\partial}{\partial X_1^a}
\frac{\partial}{\partial X_2^c}
\kappa \phi_0(Z_2)\phi_0(Z_1)
\frac{-1}{4\pi} \frac{1}{|\vec{X}_1-\vec{X}_2|} \ .
\end{align}
For spin 1/2 baryons, the trace part can be easily evaluated as 
\begin{eqnarray}
{\rm tr}
\left(
\tau^b \ba_1\tau^a
(\ba_1)^{-1}
\right)
{\rm tr}
\left(
\tau^b \ba_2\tau^c
(\ba_2)^{-1}
\right)
= \frac{64}{9}
(I_1^b I_2^b) J_1^a  J_2^c\ ,
\end{eqnarray}
where $J_i$ and $I_i$ are spin and isospin
operators, respectively. 
The potential energy is then
\begin{align}
 V
&
 \simeq -\kappa \pi^2 
\left\langle\frac{(\rho_1)^2}{k(Z_1)}\right\rangle_{(1)}
\left\langle\frac{(\rho_2)^2}{k(Z_2)}\right\rangle_{(2)}
\frac{64}{9}
(I_1^b I_2^b) J_1^a  J_2^c
\frac{\partial}{\partial X_1^a}
\frac{\partial}{\partial X_2^c}
\frac{1}{|\vec{X}_1-\vec{X}_2|} \nn\\
&
= \frac{16\kappa \pi^2}{9}
\left\langle\frac{(\rho_1)^2}{k(Z_1)}\right\rangle_{(1)}
\left\langle\frac{(\rho_2)^2}{k(Z_2)}\right\rangle_{(2)}
(I_1^b I_2^b)S_{12}
\frac{1}{|\vec r|^3}\ ,
\label{Vlarge}
\end{align}
where $S_{12}$ is defined in \eqref{S12}.
Here, we have used the relation
\begin{eqnarray}
\frac{\partial}{\partial X_1^a}
\frac{\partial}{\partial X_2^c}
\frac{1}{|\vec{X}_1-\vec{X}_2|}
=
\left(
\frac{\delta^{ac}}{3}-\hat{r}^a \hat{r}^c
\right)
\frac{3}{|\vec{r}|^3}
\end{eqnarray}
with $\vec{r}\equiv \vec{X}_1-\vec{X}_2$ and
 $\hat\vec{\!r}\equiv \vec r/|\vec r|$.

This expression can be compared with the well-known one-pion-exchange
potential
\begin{align}
V^{(\pi)}
&=\frac{1}{\pi}\left(\frac{g_A}{f_\pi}\right)^2
(I_1^bI_2^b) J_1^aJ_2^c
\frac{\del}{\del r^a}
\frac{\del}{\del r^c}
\frac{e^{-m_\pi |\vec r|}}{|\vec r|}\nn\\
&=\frac{1}{\pi}\left(\frac{g_A}{f_\pi}\right)^2
(I_1^bI_2^b)\left(
\frac{m_\pi^2}{3}(J_1^aJ_2^a)+\frac{S_{12}}{4}\left(
\frac{m_\pi^2}{3}+\frac{m_\pi}{|\vec r|}+\frac{1}{|\vec r|^2}
\right)\right)
\frac{e^{-m_\pi |\vec r|}}{|\vec r|}\ .
\label{OPEP}
\end{align}
In the chiral limit $m_\pi\ra 0$, only the tensor force remains,
and it agrees with \eqref{Vlarge} when
\begin{eqnarray}
\frac{g_A}{f_\pi}=\frac{8\pi\sqrt{\kappa\pi}}{3}\vev{\frac{\rho^2}{k(Z)}}\ ,
\end{eqnarray}
which is exactly the relation found in Ref.~\citen{HSS}.

If we use the classical values for the above expectation values,
\begin{eqnarray}
 \left\langle\frac{(\rho_1)^2}{k(Z_1)}\right\rangle_{(1)}
=\left\langle\frac{(\rho_2)^2}{k(Z_2)}\right\rangle_{(2)}
\simeq~ \rho_{\rm cl}^2 = \frac{1}{8\pi^2 a \lambda}\sqrt{\frac{6}{5}} \ ,
\end{eqnarray}
then
the potential \eqref{Vlarge} becomes
\begin{eqnarray}
 V  \simeq
-\frac{2N_c}{15 \pi^2 a \lambda}
(I_1^b I_2^b)  J_1^aJ_2^c
\frac{\partial}{\partial X_1^a}
\frac{\partial}{\partial X_2^c}
\frac{1}{|\vec{X}_1-\vec{X}_2|}
=
\frac{N_c}{30\pi^2 a \lambda}
(I_1^b I_2^b) S_{12}\frac{1}{|\vec r|^3}
 \ .
\end{eqnarray}
In a quantum evaluation for the expectation values, we will have roughly 
$\times (1.05)^2$ times the classical value above, after substituting
the numerical values \cite{HSS}.

\subsection{Intermediate distance behavior}
\label{sec:5-3}

As the baryons approach each other from asymptotics, there appear
effects of the massive meson exchange. 
At this intermediate distance, the potential (\ref{hamilf})
becomes
\begin{align}
V \simeq 
&~\kappa^2
\left[
-\frac{1}{4a^2\lambda^2}
\sum_{n=1,{\rm odd}}^\infty
\psi_n(Z_2) \psi_n(Z_1)  Y_n(|\vec{r}|)
\right.
\nonumber \\
&
+ 
\frac{256 \pi^4}{9}
(\rho_1)^2(\rho_2)^2
(I_1^bI_2^b)  J_1^aJ_2^c 
\nonumber \\
&
\hspace{10mm}\times 
\left(
\epsilon^{iaj}\epsilon^{ick}
\frac{\partial}{\partial X_1^j}
\frac{\partial}{\partial X_2^k}
+ \delta^{ca}
\frac{\partial}{\partial Z_1}
\frac{\partial}{\partial Z_2}
\right)
\sum_{n=1}^\infty
\psi_n(Z_2) \psi_n(Z_1)  Y_n(|\vec{r}|)
\nonumber \\
&
+ \frac{256 \pi^4}{9}
(\rho_1)^2(\rho_2)^2
(I_1^bI_2^b)  J_1^aJ_2^c 
\left.
\frac{\partial}{\partial X_1^a}
\frac{\partial}{\partial X_2^c}
\sum_{n=0, {\rm even}}^\infty
\phi_n(Z_2) \phi_n(Z_1)  Y_n(|\vec{r}|)
\right] \ ,
\end{align}
where we have used (\ref{green:GH})
and also dropped 
the terms including $(\partial/\partial X)(\partial/\partial Z)$,
$\psi_{2k}(Z)$, and $\phi_{2k-1}(Z)$ with $k=1,2,\cdots$,
because they vanish when they are evaluated with baryon
wavefunction, owing to the parity property of the $\psi_n(Z)$ functions.
The potential can be summarized to the following form: 
\begin{eqnarray}
V \simeq 
&&
~ \kappa^2
\left[
\left(
\frac{-1}{4a^2\lambda^2}
+ 
\frac{256 \pi^4}{9}
(\rho_1)^2(\rho_2)^2
(I_1^bI_2^b)  J_1^a J_2^c 
\epsilon^{iaj}\epsilon^{ick}
\frac{\partial}{\partial X_1^j}
\frac{\partial}{\partial X_2^k}
\right)
\sum_{n=1,{\rm odd}}^\infty
\!\!\!
\psi_n(Z_2) \psi_n(Z_1)  Y_n(|\vec{r}|)
\right.
\nonumber \\
&&
+ 
\frac{256 \pi^4}{9}
(\rho_1)^2(\rho_2)^2
(I_1^bI_2^b)  J_1^a J_2^c 
\nonumber \\
&&
\hspace{-5mm} \times 
\left(
\frac{\partial}{\partial X_1^a}
\frac{\partial}{\partial X_2^c}
\sum_{n=0, {\rm even}}^\infty
\phi_n(Z_2) \phi_n(Z_1)  Y_n(|\vec{r}|)
\right.
+
\left.
\left.
\delta^{ca}\!\!\!\!\!\!
\sum_{n=2, {\rm even}}^\infty
\lambda_n\phi_n(Z_2) \phi_n(Z_1)  Y_n(|\vec{r}|)
\right)
\right]
\ . 
\end{eqnarray}
The first term in the first line, $1/(4a^2\lambda^2)$, corresponds to
the contribution of the vector mesons that appear from the
$U(1)$ part of the gauge field. Among them, the $\omega$ meson exchange 
is at the lowest order $(n=1)$. The second term in the first
line is the contribution of the vector meson in the $SU(2)$ part of
the gauge field, whose lowest order $(n=1)$ corresponds to the $\rho$ meson.
The third line gives the contribution of the pion ($n=0$) and
axial-vector mesons ($n\ge 1$), whose lowest order is the $a_1$ meson,
in the $SU(2)$ part of the gauge field.
Note that the contributions from the
$U(1)$ part of the axial-vector and pseudo-scalar mesons are 
subleading
in the $1/N_c$ expansion.

To divide this expression into the central force and tensor force, we
use the following formula for the Yukawa potential $Y_n(|\vec{r}|)$, 
\begin{eqnarray}
 \frac{\partial}{\partial X_1^a}
\frac{\partial}{\partial X_2^c}
Y_n(|\vec{r}|)
=
\left(
\frac{\delta^{ac}}{3}-\hat{r}^a \hat{r}^c
\right)
\left(
\frac{3}{|\vec{r}|^2} + \frac{3\sqrt{\lambda_n}}{|\vec{r}|} + \lambda_n
\right)Y_n(|\vec{r}|)-\frac{\delta^{ab}}{3}\lambda_n Y_n(|\vec r|)\ .
\end{eqnarray}
Then, we obtain the potential energy due to the central force and
tensor force, 
\begin{align}
V = 
&\,
V_{\rm C} + S_{12}V_{\rm T} \ , 
\\
V_{\rm C} 
= 
&\,
\kappa^2
\left[
-\frac{1}{4a^2\lambda^2}
\sum_{n=1,{\rm odd}}^\infty
\psi_n(Z_2) \psi_n(Z_1)  Y_n(|\vec{r}|)
\right.
\nonumber \\
&
+ \frac{256\pi^4}{9}
(\rho_1)^2(\rho_2)^2
(I_1^b I_2^b)(J_1^a J_2^a)
\nonumber \\
&
\times
\left.\frac{2}{3}
\left(-\!\!\!\!
\sum_{n=1,{\rm odd}}^\infty\lambda_n
\psi_n(Z_2) \psi_n(Z_1)  Y_n(|\vec{r}|)
+\!\!\!\!
\sum_{n=2,{\rm even}}^\infty\lambda_n
\phi_n(Z_2) \phi_n(Z_1)  Y_n(|\vec{r}|)
\right)
\right] \ , 
\label{VC}
\\
V_{\rm T} 
=
&\,
\kappa^2
\frac{64\pi^4}{27}
(\rho_1)^2(\rho_2)^2
(I_1^b I_2^b )
\left(
\sum_{n=1,{\rm odd}}^\infty
\psi_n(Z_2) \psi_n(Z_1) 
\left(\frac{3}{|\vec{r}|^2}+\frac{3\sqrt{\lambda_n}}{|\vec{r}|}
+\lambda_n\right)
Y_n(|\vec{r}|)
\right.
\nonumber \\
&
\hspace{20mm}
\left.
-
\sum_{n=0,{\rm even}}^\infty
\phi_n(Z_2) \phi_n(Z_1) 
\left(\frac{3}{|\vec{r}|^2}+\frac{3\sqrt{\lambda_n}}{|\vec{r}|}
+\lambda_n\right)
Y_n(|\vec{r}|)
\right)
\ .
\label{VT}
\end{align}
We note that
these expressions can be reproduced from the leading order of
the one-boson-exchange potential computed
from tree-level Feynman diagram with the nucleon-nucleon-meson couplings
obtained in Ref.~\citen{HSS}. 
See Appendix \ref{EPr} for detail.

Let us look at the contribution from light mesons. The $\omega$ meson gives
only the central force, the first line in $V_{\rm C}$ with $n=1$. It is
\begin{eqnarray}
 V_{\rm C}^{(\omega)} = -\kappa^2 
\frac{1}{4a^2 \lambda^2}
\langle\psi_1(Z_1)\rangle_{(1)} 
\langle\psi_1(Z_2)\rangle_{(2)} 
Y_1(|\vec{r}|) \ .
\end{eqnarray}
For the potential between the same types of baryons, this expression of
course reduces to
\begin{eqnarray}
 V_{\rm C}^{(\omega)} = -\kappa^2 
\frac{1}{4a^2 \lambda^2}
\langle\psi_1(Z)\rangle^2 
Y_1(|\vec{r}|) \ . 
\end{eqnarray}
Since the Yukawa potential $Y_n$ defined in \eqref{Yn} is negative,
the $\omega$ meson exchange gives a strong repulsion force
in the central force.
The $\omega$ meson is the lightest vector meson that comes
from the $U(1)$ part of the five-dimensional gauge field. The
instanton is electrically charged under this $U(1)$, 
so the baryons should
have this universal repulsive force. The $\omega$ meson exchange
manifests its lowest term in the KK decomposition of the
higher-dimensional ``electric'' repulsion.

The $\rho$ meson exchange can be seen in $SU(2)$ components of $n=1$. 
The central force is
\begin{eqnarray}
  V_{\rm C}^{(\rho)} = \kappa^2 
\frac{512\pi^4}{27}
(\rho_1)^2(\rho_2)^2
(I_1^b I_2^b)(J_1^a J_2^a )\,
\psi_1(Z_2) \psi_1(Z_1) (-\lambda_1) Y_n(|\vec{r}|) \ ,
\end{eqnarray}
while the tensor force is 
\begin{eqnarray}
 V_{\rm T}^{(\rho)} = \kappa^2 
\frac{64\pi^4}{27}
(\rho_1)^2(\rho_2)^2
(I_1^b I_2^b )\,
\psi_1(Z_2) \psi_1(Z_1) 
\left(\frac{3}{|\vec{r}|^2}\!+\!\frac{3\sqrt{\lambda_1}}{|\vec{r}|}
\!+\!\lambda_1\!\!\right)\!
Y_1(|\vec{r}|) \ .
\nonumber 
\end{eqnarray}
When two baryons are of the same type, these reduce to
\begin{eqnarray}
&&  V_{\rm C}^{(\rho)} = \kappa^2 
\frac{512\pi^4}{27}
\langle\rho^2\rangle^2
(I_1^b I_2^b)(J_1^a J_2^a )\,
\langle\psi_1(Z)\rangle^2  (-\lambda_1) Y_1(|\vec{r}|) \ ,
\nn \\
&&
  V_{\rm T}^{(\rho)} = \kappa^2 
\frac{64\pi^4}{27}
\langle\rho^2\rangle^2
(I_1^b I_2^b )\,
\langle\psi_1(Z)\rangle^2
\left(\frac{3}{|\vec{r}|^2}\!+\!\frac{3\sqrt{\lambda_1}}{|\vec{r}|}
\!+\!\lambda_1\!\!\right)\!
Y_1(|\vec{r}|) \ .
 \end{eqnarray}

The strength of this tensor force can be compared with the strength of
the pion tensor force (\ref{Vlarge}).
The ratio of the front coefficients is given by 
${-\langle \psi_1(Z)\rangle^2}/{\langle\phi_0(Z)\rangle^2}$.
The classical evaluation of this gives
\begin{eqnarray}
  \frac{-\langle \psi_1(Z)\rangle^2}{\langle\phi_0(Z)\rangle^2}
= -\kappa \pi \langle \psi_1(Z)\rangle^2
\simeq -\pi (0.597)^2 \sim -1\ .
\end{eqnarray}
Let us see how this $\rho$ meson exchange may change the result of the
$\pi$ exchange. At the length scale $r\sim 1~\mbox{fm}$ $\sim
200~\mbox{MeV}^{-1}$, the pion behaves as a massless particle while the
$\rho$ meson is massive. The above ratio denotes that the coefficient is
of the same order, while the sign is opposite to each other. The
normalization of the Yukawa potential gives the following rough ratio
\begin{eqnarray}
\frac{ e^{-m_\rho |\vec{r}|}}{e^{-m_\pi |\vec{r}|}} \sim {\cal O}(0.1) \ .
\end{eqnarray}
This means that the $\rho$ meson exchange does not contribute much to
the tensor force at
the distance scale $\sim 1 ~\mbox{fm}$.

\section{Summary}
\label{sec:6}

In this paper, we have deduced the nuclear force at short distance in
large $N_c$ strongly coupled QCD, by applying gauge/string duality.

In the D4-D8 model \cite{SaSu1,SaSu2} of holographic QCD, baryons are 
instantons in (1+4)-dimensional YMCS theory. We have explicitly
constructed
a two-instanton solution in the theory 
by employing ADHM construction of instantons. The analytic expression for
the potential energy plus kinetic terms of the baryon, {\it i.e.},~the
effective Hamiltonian of quantum mechanics for two-baryon system,
has been derived, for the distance 
${\cal O}(1/(\sqrt{\lambda} M_{\rm KK})) < r 
< {\cal O}(1/M_{\rm KK})$.

The evaluation of this interaction Hamiltonian for specific two-nucleon 
states, labeled by spin $J_i^3$ and isospin $I_i^3$ with $i=1,2$, 
provides the nuclear force at the short distance scale.
We have obtained a central force (\ref{fresult0}) 
as well as a tensor force (\ref{fresult1}).
The central force exhibits a strong repulsive core of a nucleon.
As the repulsive core has been mysterious from the viewpoint of
strongly coupled QCD, our result is of importance as a derivation of the
repulsive core from the ``first principle'' of QCD, in the large
$N_c$ expansion and also with the gauge/string duality. 

The obtained nucleon-nucleon 
potential at short distances has $r^{-2}$ behavior.
Technically speaking, this comes from the harmonic potential in
four spatial dimensions including the holographic extra dimension.
It would be interesting to fit this peculiar behavior with the
experimental observation or the recent lattice result
\cite{Ishii:2006ec}. 

As our result is for the short distance, it is important to generalize
our analysis to a larger distance scale. For $r>{\cal O}(1/M_{\rm KK})$,
the effect of the curvature along $z$ becomes indispensable, thus, one
needs to construct a two-instanton solution in curved space.
This may lead to an analysis of a deuteron system in holographic QCD.
For this, the inclusion of quark mass to the model 
\cite{Aharony:2008an} that should
make the pion massive may be important.
On the other hand, the height of the nucleon-nucleon potential
at $r=0$ is of interest, but its derivation has turned out to be
difficult, as we explained in \S\ref{sec:2-2}. 
Further effort along these 
directions may reveal some more interesting physics in QCD, via the 
holography.

\section*{Acknowledgements}

We would like to thank H.~Fujii, T.~Hatsuda, K.~Itakura, 
T.~Matsui, T.~Nakatsukasa, M.~Nitta, S.~Sasaki,
and Y.~Yamaguchi for valuable 
discussions.
The work of K.H. and T.S. is partially supported by 
a Grant-in-Aid for Young Scientists (B), MEXT, Japan.
The work of S.S. is supported in part by
JSPS Grant-in-Aid for Creative Scientific Research
No. 19GS0219 and also by World Premier International
Research Center Initiative (WPI Initiative), MEXT, Japan.
We would like to thank 
the Yukawa Institute for Theoretical Physics at Kyoto University,
where we discussed this topic during the workshop YITP-W-08-04 on 
``Development of Quantum Field Theory and String Theory''.

\appendix
%

\section{Notation for quaternion}
\label{appA}
A quaternion $\bq\in\bH$ is given as a linear combination of
 the form
\begin{eqnarray}
\bq=q^4-q^1 I-q^2 J-q^3 K\ ,~~~(q^m\in\bR\ ,~~m=1,2,3,4)
\end{eqnarray}
where $I$, $J$, and $K$ satisfy
\begin{eqnarray}
I^2=J^2=K^2=-1\ ,~~IJ=-JI=K\ ,~
JK=-KJ=I\ ,~KI=-IK=J\ .
\nn\\
\label{IJK}
\end{eqnarray}
Since $(-i\tau^1,-i\tau^2,-i\tau^3)$ satisfy the same algebra as
$(I,J,K)$, a quaternion can be represented as a $2\times 2$ 
complex matrix of the form
\begin{eqnarray}
\bq=q^4+iq^a\tau^a\ .
\label{cpxrep}
\end{eqnarray}

The conjugate and norm of a quaternion $\bq$ are defined as
$\bq^\dag =q^4+q^1 I+q^2 J+q^3 K$ and
$|\bq|\equiv \sqrt{\bq^\dag \bq} =\sqrt{\bq \bq^\dag}=\sqrt{q^mq^m}$\ ,
respectively.

The product of two quaternions $\bq,\bw\in\bH$ follows from the relation
(\ref{IJK}) and it can be written in terms of the $2\times 2$ complex
matrix representation (\ref{cpxrep})
\begin{eqnarray}
 \bq\bw=q^4w^4-\vec q\cdot\vec w+
i(q^4\vec w+w^4\vec q-\vec q\times\vec w)\cdot \vec\tau\ ,
\end{eqnarray}
where $\vec q=(q^1,q^2,q^3)$, etc.
We also use the following notation:
\begin{eqnarray}
&&(\bq\cdot \bw)\equiv \half(\bq^\dag\bw+\bw^\dag\bq)
=q^m w^m \ ,\\
&&(\bq\times\bw)\equiv \half(\bq^\dag\bw-\bw^\dag\bq)
=i(q^4\vec w-w^4\vec q+\vec q\times\vec w)\cdot\vec\tau \ .
\end{eqnarray}

An element of $Sp(n)$ is defined as a $n\times n$
quaternionic matrix $Q$ satisfying $Q^\dag Q = 1_n$.\footnote{
$Sp(n)$ in this paper is the unitary symplectic group,
which is also written as $USp(2n)$.}
In particular, an element of $Sp(1)$ is a quaternion
satisfying
\begin{eqnarray}
\bq^\dag \bq=q^mq^m=1\ ,
\end{eqnarray}
which is equivalent to the condition for an element of $SU(2)$
in the $2\times 2$ complex matrix representation \eqref{cpxrep}.

\section{ADHM construction}
\label{appB}
\subsection{ADHM construction for $Sp(n)$ instantons}
Here, we briefly review the ADHM construction \cite{ADHM}.
(See for example Ref.~\citen{ADHMreview} for a review.)
Since $SU(2)=Sp(1)$, the ADHM construction for the $Sp(n)$ instantons
is useful for our purpose.

We define an $(n+k)\times k$ quaternionic
matrix of the form
\begin{eqnarray}
\Delta(x)=a+b\,\bx\ ,
\end{eqnarray}
where $a$ and $b$ are
$(n+k)\times k$ quaternionic matrices and
$\bx=x^4-x^1I-x^2J-x^3K\in\bH$ is a quaternion
composed of the coordinate of the four-dimensional space
$(x^1,x^2,x^3,x^4)=(\vec x,z)$.
The matrices $a$ and $b$ are required to satisfy
that $a^\dag a$ and $b^\dag b$ are $k\times k$ real
symmetric matrices, and $a^\dag b$
is a $k\times k$ symmetric quaternionic matrix.
These conditions for the matrices $a$ and $b$
are equivalent to the constraint that
the matrix $\Delta$ satisfies 
\begin{eqnarray}
\Delta^\dag\Delta=L(x)
\label{constSp}
\end{eqnarray}
with a $k\times k$ real symmetric matrix $L(x)$.
We also implicitly assume that the matrices $a$ and $b$
are generic and they are matrices of rank $k$.

The ADHM gauge field for the $k$-instanton configuration
is given by
\begin{eqnarray}
A_m(x)=-iU(x)^\dag \del_m U(x)\ ,
\label{gaugeSp}
\end{eqnarray}
where $U(x)$ is $(n+k)\times n$ quaternionic matrix satisfying
\begin{eqnarray}
\Delta^\dag U=0\ ,~~U^\dag U=1_n\ .
\label{U}
\end{eqnarray}
Note that the matrix $U(x)$ is defined up to a transformation
\begin{eqnarray}
 U(x)\rightarrow U(x) g(x)\ , ~~~(g(x)\in Sp(n))\ ,
\end{eqnarray}
which acts as a gauge transformation for the gauge field (\ref{gaugeSp}).

The gauge field (\ref{gaugeSp})
as well as the constraint (\ref{constSp})
 is invariant under
\begin{eqnarray}
\Delta(x) \ra Q\Delta(x) R\ ,~~L(x)\ra R^T L(x)R\ ,
\label{sptr}
\end{eqnarray}
where $Q\in Sp(n+k)$ and $R\in GL(k,\bR)$.
By using this transformation, the matrix $b$ can be fixed as
\begin{align}
b=\left(
\begin{array}{c}
0\\
-1_k
\end{array}
\right)
\ ,
\label{canonical-b}
\end{align}
and then $\Delta$ is of the canonical form
\begin{eqnarray}
\Delta(x)=\left(
\begin{array}{c}
Y\\
X-\bx\, 1_k
\end{array}
\right)
\ ,~~
\label{canonical}
\end{eqnarray}
where $X$ is a $k\times k$ symmetric quaternionic matrix
 and $Y$ is an $n\times k$ quaternionic matrix
 such that $Y^\dag Y+X^\dag X$ is a $k\times k$ symmetric real matrix.
Note that we have not completely used the transformation \eqref{sptr}.
In fact, a transformation \eqref{sptr} with $R\in O(k)$ and
\begin{eqnarray}
Q=\mat{q,,,R^T}\ ,~~~(q\in Sp(n))
\label{O2}
\end{eqnarray}
leaves \eqref{canonical-b} invariant.

\subsection{$Sp(1)=SU(2)$  one-instanton}
As an exercise, let us consider the $n=k=1$ case.
Using the canonical form \eqref{canonical}, we have
\begin{eqnarray}
\Delta(x)=\left(
\begin{array}{c}
\by\\
\bX-\bx
\end{array}
\right)
\ ,
\label{D1}
\end{eqnarray}
with $\by,\bx,\bX\in \bH$.
In this case, \eqref{constSp} is satisfied
without imposing further constraints.

The condition \eqref{U} is solved by
\begin{eqnarray}
U^\dag=
\frac{1}{\sqrt{\xi^2+\rho^2}}\left(\by(\bx-\bX)\by^{-1},\by
\right)\ ,
\label{Udag}
\end{eqnarray}
where
$\xi\equiv\sqrt{|\bx-\bX|^2}$ and $\rho\equiv \sqrt{|\by|^2}$.

Then the ADHM gauge field (\ref{gaugeSp}) is
\begin{eqnarray}
A_m=-i\ba(f(\xi)\,g\del_m g^{-1})\ba^{-1}\ ,
\end{eqnarray}
where
\begin{eqnarray}
f(\xi)=\frac{\xi^2}{\xi^2+\rho^2}\ ,~~g=\frac{\bx-\bX}{\xi}\ ,
\end{eqnarray}
and $\ba\equiv \by/\rho$ is an element of $Sp(1)=SU(2)$.

This is the BPST instanton solution.
The moduli parameters $\bX$, $\rho$, and $\ba$ correspond to
 the position, size, and gauge orientation of the instanton.
The $\ba$-dependence of the gauge field can be
eliminated by a global gauge transformation. However, it is known
that this degree of freedom is also physically relevant when
we quantize the system via moduli space approximation method.

\subsection{$Sp(1)=SU(2)$ two-instanton}

For $n=1$ and $k=2$, the ansatz (\ref{canonical}) can be written as
\begin{align}
 Y=(\by_1,\by_2)\ ,~X=\mat{\bX_1,\bw,\bw,\bX_2}\ ,~
\Delta(x)=\left(
\begin{array}{cc}
\by_1&\by_2\\
\bX_1-\bx&\bw\\
\bw&\bX_2-\bx
\end{array}
\right)\ .
\label{n1k2}
\end{align}
The constraint (\ref{constSp}) requires
\begin{eqnarray}
Y^\dag Y+X^\dag X=\mat{|\by_1|^2+|\bX_1|^2+|\bw|^2,
\by_1^\dag \by_2+\bX_1^\dag \bw+\bw^\dag\bX_2,
\by_2^\dag \by_1+\bX_2^\dag \bw+\bw^\dag\bX_1,
|\by_2|^2+|\bX_2|^2+|\bw|^2}
\end{eqnarray}
to be a real symmetric matrix and hence
\begin{eqnarray}
\by_1^\dag \by_2-\by_2^\dag\by_1+\br^\dag\bw-\bw^\dag\br
=0\ ,
\label{yyrw}
\end{eqnarray}
where we have defined $\br=\bX_1-\bX_2$.
This equation is solved when
\begin{eqnarray}
\bw=\frac{\br}{|\br|^2}(\by_2\times \by_1)
+\alpha\,\br\ ,
\label{w}
\end{eqnarray}
with $\alpha\in\bR$.
This parameter $\alpha$ can be eliminated by
the residual $O(2)$ symmetry in (\ref{O2}) with $q=1$ and $R\in O(2)$
\cite{DoKhMa}.
We will set $\alpha=0$ in this paper.

After all, we have 4 quaternionic parameters $\by_1$, $\by_2$,
$\bX_1$, and $\bX_2$ to parameterize the two instanton moduli space.
It can be shown that when the separation of the two instantons
is sufficiently large, the two-instanton configuration can be approximated
by the superposition of two 1-instanton configurations.
Then, $\bX_i$ ($i=1,2$) represents the positions
of the two instantons,
 $\rho_i\equiv\sqrt{|\by_i|^2}$ corresponds to
their size, and $\ba_i\equiv \by_i/\rho_i$ is their $SU(2)$ orientation.
This fact can be explicitly seen in the effective Hamiltonian
(\ref{Ham2}).

\section{Evaluation of $H_{\rm pot}^{(SU(2))}$}
\label{app:SU2}

In this section, we compute
\begin{eqnarray}
\int d^4x\, z^2 \Tr F_{mn}^2\ ,
\end{eqnarray}
where $z=x^4$. 
We will follow the strategy of Ref.~\citen{Ma},
in which
\begin{eqnarray}
\int d^4x\, |\bx|^2 \Tr F_{mn}^2\ ,
\end{eqnarray}
is calculated.

To evaluate $\Tr F_{mn}^2$,
we use the useful formula \cite{Os2}
\begin{eqnarray}
\tr F_{mn}^2=\half\epsilon^{mnpq}\tr F_{mn}F_{pq}
=-\Box\Box\log\det L\ , 
\label{trF2}
\end{eqnarray}
where $\Box\equiv\del_m\del_m$ and $L(x)$ is defined in \eqref{constSp}.

In general, $L(x)$ can be written as
\begin{eqnarray}
L(x)=\Lambda |\bx|^2-2\gamma_m x^m + A\ ,
\end{eqnarray}
where $\Lambda=b^\dag b$ and $A=a^\dag a$ are positive definite
 $k\times k$ real symmetric matrices, and
$\gamma_m$ ($m=1,2,3,4$) are  $k\times k$ real symmetric matrices.
 For the canonical form \eqref{canonical},
they are given by
\begin{eqnarray}
 \Lambda=1\ ,~~\gamma_m= X_m \ ,~~A=Y^\dag Y+X^\dag X\ ,
\label{LgA}
\end{eqnarray}
where $X_m$ is the $k\times k$ real symmetric matrix satisfying
 $X=X_4-X_1 I- X_2 J- X_3 K$.

The result we are going to prove is
\begin{eqnarray}
\int d^4x\, z^2 \tr F_{mn}^2
=8\pi^2\tr\left[(\gamma_4\Lambda^{-1})^2
-(\gamma_1\Lambda^{-1})^2-(\gamma_2\Lambda^{-1})^2-(\gamma_3\Lambda^{-1})^2
+A\Lambda^{-1}\right]\ .
\label{result0}
\end{eqnarray}
For the canonical form with \eqref{LgA}, we obtain
\begin{eqnarray}
\int d^4x\, z^2 \tr F_{mn}^2
=8\pi^2\tr\left(2(X_4)^2+Y^\dag Y\right)\ .
\label{result1}
\end{eqnarray}
Note that this formula implies
\begin{eqnarray}
\int d^4x\, |\bx|^2 \tr F_{mn}^2
=16\pi^2\tr\left(2Y^\dag Y+X^\dag X\right)\ ,
\label{pot}
\end{eqnarray}
which reproduces the result in Ref.~\citen{Ma}.

To show \eqref{result0}, we can set $\Lambda=1$ without loss of
generality. The $\Lambda$ dependence can easily be recovered by
the transformation \eqref{sptr}.
Integrating by parts, we obtain
\begin{align}
&\int d^4 x\, z^2\tr F_{mn}^2\nn\\
&=
-\lim_{R\ra\infty}\int_{S^3}d^3\Omega\, R^2 x^m
\Big[
z^2\del_m \Box\log\det L
- 2z\delta_{m4}\, \Box\log\det L
+ 2 \del_m\log\det L\Big]\ ,
\label{intpart}
\end{align}
where $R^2=|\bx|^2$ and $d^3\Omega$ is the volume element
of unit $S^3$.
Here, we have used Gauss's law
\begin{eqnarray}
\int d^4 x\,\del_m W_m=
\lim_{R\ra\infty}\int_{S^3} d^3\Omega\,R^3 n_m W_m\ ,
\end{eqnarray}
where $n_m=x^m/R$ is a unit normal vector of the $S^3$.

We can easily check the following formulas
\begin{eqnarray}
\del_m\log\det L&=&\tr(V_m)\ ,\\
\Box\log\det L&=&8\tr(L^{-1})-\tr(V_m^2)\ ,\\
\del_m\Box\log\det L&=&-12\tr(L^{-1}V_m)+2\tr(V_mV_n^2)\ ,
\end{eqnarray}
where $V_m\equiv L^{-1}\del_m L$.
Using these, \eqref{intpart} becomes
\begin{align}
&\int d^4 x\, z^2\tr F_{mn}^2\nn\\
&=
-\lim_{R\ra\infty}\int_{S^3}d^3\Omega\, R^2
\tr\Big[
-12z^2L^{-1}x^mV_m+2z^2\, x^mV_m V_n^2
\nn\\
& ~~~~~~~~~~~~~~~~~~~~~~~~~~~~~~~~~~~~~~~~~~~~~~~
-16 z^2 L^{-1}
+2 z^2\,V_n^2+2\, x^mV_m
\Big] \ .
\end{align}
Inserting the relation
$V_m=L^{-1}\del_m L=2L^{-1}(x^m-\gamma_m)$, we obtain
\begin{align}
&\int d^4 x\, z^2\tr F_{mn}^2\nn\\
=&
-\lim_{R\ra\infty}\int_{S^3}d^3\Omega\, R^2
\tr\Big[4(R^2-4z^2)L^{-1}
-4\,x^m\gamma_mL^{-1}
-24z^2\,(R^2-x^m\gamma_m)L^{-2}\nn\\
&+8z^2\left(L+2(R^2-x^m\gamma_m)\right)
\left(R^2L^{-2}+L^{-1}\gamma_mL^{-1}\gamma_m-
2L^{-2}x^m\gamma_m\right)L^{-1}
\Big]\ .
\end{align}
We are only interested in the $\cO(R^{0})$ terms
in the integrand of the right-hand side of
this equation.
Then, recalling $L=\cO(R^2)$, we obtain
\begin{eqnarray}
\int d^4 x\, z^2\tr F_{mn}^2
&=&
-\lim_{R\ra\infty}\int_{S^3}d^3\Omega\, 
\tr\Big[P_1+P_2+P_3+P_4+P_5\Big]\ ,
\end{eqnarray}
where
\begin{align}
&
P_1=R^6\cdot 4L^{-3}(R^2-4z^2)\ , \quad
P_2=R^4\cdot 4x^n\gamma_n\,L^{-3}(-5R^2+14z^2) \ , \quad
\nn\\
&
P_3 = R^{-4}\cdot 16(x^n\gamma_n)^2(2R^2-3z^2) \ , \quad
P_4 = R^{-2}\cdot 8A(R^2-6z^2) \ , \quad
\nn\\
&
P_5 = R^{-2}\cdot 24\gamma_n^2z^2 \ . \qquad
\label{1-5}
\end{align}
Useful formulas to evaluate the integral are
\begin{align}
\int_{S^3}d^3\Omega=&\,
2\pi^2\ ,\nn\\
\int_{S^3}d^3\Omega\,x^mx^n=&\,
\frac{\pi^2}{2}R^2\delta_{mn}\ ,\nn\\
\int_{S^3}d^3\Omega\,(x^m)^2 (x^n)^2=&\,
\frac{\pi^2}{12}R^4(1+2\delta_{mn})\ .~~~(\mbox{no sum for $m,n$})
\label{formula}
\end{align}

To evaluate the integral of $P_1$ in \eqref{1-5}, we expand $L^{-3}$
\begin{eqnarray}
L^{-3}=\frac{1}{R^6}
\left(1+6\frac{x^m\gamma_m}{R^2}-3\frac{A}{R^2}
+24\frac{(x^m\gamma_m)^2}{R^4}+\cO(R^{-3})
\right)
\end{eqnarray}
and
\begin{eqnarray}
-\lim_{R\ra\infty}\int_{S^3}d^3\Omega\, \tr P_1 
&=&
-\lim_{R\ra\infty}\int_{S^3}d^3\Omega\, 
\tr\Big[
R^6\,4L^{-3}(R^2-4z^2)\Big]\nn\\
&=&
-\lim_{R\ra\infty}\int_{S^3}d^3\Omega\, 
\tr\Bigg[
\,4(R^2-4z^2)
+4(R^2-4z^2)\cdot 6\frac{x^m\gamma_m}{R^2}\nn\\
&&~~~~~+4(R^2-4z^2)\left(
-\frac{3}{R^2}A+\frac{24}{R^4}(x^m\gamma_m)^2
\right)
\Bigg] \ .
\end{eqnarray}
Using the formulas \eqref{formula}, we see that
the first term vanishes. The second term also vanishes
since it is odd under $x\ra -x$. From the third term,
we obtain
\begin{eqnarray}
-\lim_{R\ra\infty}\int_{S^3}d^3\Omega\, \tr P_1 
&=&
-16\pi^2\tr(\gamma_1^2+\gamma_2^2+\gamma_3^2
-3\gamma_4^2)\ .
\end{eqnarray}
The integrals for the other $P_2, \cdots, P_5$ 
are calculated in a similar manner
The results are
\begin{eqnarray}
-\lim_{R\ra\infty}\int_{S^3}d^3\Omega\, \tr P_2
&=&
8\pi^2\tr\left(4(\gamma_1^2+\gamma_2^2+\gamma_3^2)
-3\gamma_4^2\right)\ ,\\
-\lim_{R\ra\infty}\int_{S^3}d^3\Omega\, \tr P_3
&=&
-4\pi^2\tr\left(3(\gamma_1^2+\gamma_2^2+\gamma_3^2)
+\gamma_4^2\right)\ ,\\
-\lim_{R\ra\infty}\int_{S^3}d^3\Omega\, \tr P_4
&=&\pi^2\tr(A)\ ,\\
-\lim_{R\ra\infty}\int_{S^3}d^3\Omega\, \tr P_5
&=&-12\pi^2\tr(\gamma_m^2)\ .
\end{eqnarray}
Summing up all these, we finally obtain
\begin{eqnarray}
\int d^4 x\, z^2\tr F_{mn}^2
=8\pi^2\tr\left(
-(\gamma_1^2+\gamma_2^2+\gamma_3^2)+\gamma_4^2+A
\right)\ .
\end{eqnarray}

\section{Laplacian of the two-instanton moduli space}
\label{app:metric}
Here, we outline the derivation of the expression \eqref{lap} and \eqref{lap2}.

Consider a metric given as
\begin{eqnarray}
g_{\alpha\beta}=\ol g_{\alpha\beta}+h_{\alpha\beta}\ ,
\label{h}
\end{eqnarray}
where $\ol g_{\alpha\beta}$ is a constant metric and
 $h_{\alpha\beta}\ll 1$ is a small perturbation. Then, omitting the
 $\cO(h^2)$ terms, we have
\begin{eqnarray}
g^{\alpha\beta}=\ol g^{\alpha\beta}-h^{\alpha\beta}\ ,~~
\sqrt{g}=\sqrt{\ol g}\left(1+\half h^\alpha_{~\alpha}\right)\ ,
\end{eqnarray}
\begin{eqnarray}
\nabla^2&=&\frac{1}{\sqrt{g}}\del_\alpha\sqrt{g}\,g^{\alpha\beta}\del_\beta\nn\\
&=&\nabla_0^2-h^{\alpha\beta}\del_\alpha\del_\beta+
\half(\del_\beta h^\alpha_{~\alpha})\del^\beta-(\del_\alpha
h^{\alpha\beta})\del_\beta \ ,
\label{lap0}
\end{eqnarray}
where $(\ol g^{\alpha\beta})$ is the inverse matrix of
$\ol g_{\alpha\beta}$, $\nabla_0^2=\ol g^{\alpha\beta}\del_\alpha\del_\beta$,
 $g=\det(g_{\alpha\beta})$, $\ol g=\det(\ol g_{\alpha\beta})$.
Here, raising and lowering the indices is done with 
 $\ol g^{\alpha\beta}$ and $\ol g_{\alpha\beta}$.
We apply these formulas to the metric \eqref{metric}, in which
 $ds_0^2$ and $ds_1$ correspond to $\ol g_{\alpha\beta}$ and
 $h_{\alpha\beta}$,  respectively.

A little algebra shows
\begin{align}
h^\alpha_{~\alpha}&=\frac{2}{|\br|^2}
(|\by_1|^2+|\by_2|^2)\ ,\\
(\del_\beta h^\alpha_{~\alpha})\del^\beta
&=\frac{2}{|\br|^2}\left[
\left(\by_1\cdot\frac{\del}{\del \by_1}\right)
+\left(\by_2\cdot\frac{\del}{\del \by_2}\right)
\right]+\cO(|\br|^{-3})\ ,\\
(\del_\alpha h^{\alpha\beta})\del_\beta
&=\frac{2}{|\br|^2}\left[
\left(\by_1\cdot\frac{\del}{\del\by_1}\right)
+\left(\by_2\cdot\frac{\del}{\del\by_2}\right)
\right]+\cO(|\br|^{-3})\ ,
\end{align}
\begin{align}
h^{\alpha\beta}\del_\alpha\del_\beta
=&\frac{1}{|\br|^2}\Bigg[
\frac{\rho_2^2}{2}
\left(\frac{\del}{\del\by_1}\cdot\frac{\del}{\del\by_1}\right)
+\frac{\rho_1^2}{2}
\left(\frac{\del}{\del\by_2}\cdot\frac{\del}{\del\by_2}\right)
+\left(\by_1\cdot \frac{\del}{\del\by_1}\right)
\left(\by_2\cdot \frac{\del}{\del\by_2}\right)
\nn\\
&~~~
-\left(\by_2\cdot \frac{\del}{\del\by_1}\right)^2
-\left(\by_1\cdot \frac{\del}{\del\by_2}\right)^2
-(\by_1\cdot\by_2)
\left(\frac{\del}{\del\by_1}\cdot\frac{\del}{\del\by_2}\right)
\nn\\
&~~~+\,\epsilon_{IJKL}\,y_1^I y_2^J
\frac{\del}{\del y_1^K}\frac{\del}{\del y_2^L}
+\,y_1^I y_2^J \frac{\del}{\del y_1^J}
 \frac{\del}{\del y_2^I}\,\Bigg]\ .
\end{align}
{}From these, we can easily obtain \eqref{lap}.

The following formulas are useful to obtain the expression in \eqref{lap2}: 
\begin{eqnarray}
\left(\by_i\cdot \frac{\del}{\del \by_i}\right)
=\rho_i\frac{\del}{\del\rho_i}\ ,
\end{eqnarray}
\begin{align}
\left(\frac{\del}{\del \by_i}\cdot\frac{\del}{\del\by_i}\right)
=\frac{\del^2}{\del\rho_i^2}
+\frac{3}{\rho_i}\frac{\del}{\del\rho_i}
-\frac{4}{\rho_i^2}I_i^aI_i^a
=\frac{\del^2}{\del\rho_i^2}
+\frac{3}{\rho_i}\frac{\del}{\del\rho_i}
-\frac{4}{\rho_i^2}J_i^aJ_i^a\ ,
\end{align}
\begin{align}
I_1^aI_2^a&=-\frac{1}{4}\left[
(\by_1\cdot \by_2)
\left(\frac{\del}{\del \by_1}\cdot \frac{\del}{\del\by_2}\right)
-y_1^I y_2^J\frac{\del}{\del y_1^J}\frac{\del}{\del y_2^I}
-\epsilon_{IJKL}y_1^Iy_2^J
\frac{\del}{\del y_1^K}\frac{\del}{\del y_2^L}
\right]\ ,\\
J_1^aJ_2^a&=-\frac{1}{4}\left[
(\by_1\cdot \by_2)
\left(\frac{\del}{\del \by_1}\cdot \frac{\del}{\del\by_2}\right)
-y_1^I y_2^J\frac{\del}{\del y_1^J}\frac{\del}{\del y_2^I}
+\epsilon_{IJKL}y_1^Iy_2^J
\frac{\del}{\del y_1^K}\frac{\del}{\del y_2^L}
\right]\ .
\end{align}

\section{Evaluation of $H_{\rm pot}^{(U(1))}$}
\label{app:U1}

Here, we rederive the result \eqref{HU1} in a more systematic way.
{}From the expression in \eqref{L2}, we obtain
\begin{eqnarray}
\log\det L=\log f_1+\log f_2+\log f_3\ ,
\end{eqnarray}
where
\begin{align}
f_i(x)&=\rho_i^2+|\bx-\bX_i|^2+|\bw|^2\ ,~~(i=1,2)\\
f_3(x)&=1-
\frac{e(x)^2}{f_1(x) f_2(x)}
\end{align}
with
\begin{eqnarray}
e(x)&=(\by_1\cdot\by_2) + \left(\bw\cdot(\bX_1+\bX_2-2\bx)\right)\ .
\end{eqnarray}
Substituting \eqref{A0sol} into \eqref{u1pot}, we obtain
\begin{align}
H_{\rm pot}^{(U(1))}&=
\frac{aN_c}{2}\frac{1}{(32\pi^2 a)^2}
\int d^3 x dz\,
\left(\del_M\Box(\log f_1+\log f_2+\log f_3)\right)^2\nn\\
&=\frac{aN_c}{2}\frac{1}{(32\pi^2 a)^2}
\int d^3 x dz\,\Bigg[
\sum_{i=1,2}(\del_M\Box\log f_i)^2+(\del_M\Box\log f_3)^2+\nn\\
&~~~+2(\del_M\Box\log f_1)(\del_M\Box\log f_2)
+2\sum_{i=1,2}(\del_M\Box\log f_i)(\del_M\Box\log f_3)
\Bigg]\ .
\label{HU1exp}
\end{align}
The following formulas are useful for evaluating this integral:
\begin{align}
\Box\log f_i
 &=\frac{4(|\bx-\bX_i|^2+2(\rho_i^2+|\bw|^2))}
{(|\bx-\bX_i|^2+\rho_i^2+|\bw|^2)^2}\ ,\\ 
\Box\Box\log f_i &=-\frac{96(\rho_i^2+|\bw|^2)^2}
{(|\bx-\bX_i|^2+\rho_i^2+|\bw|^2)^4}\ ,\\
\Box\Box\Box\log f_i &=-
\frac{1536(\rho_i^2+|\bw|^2)^2(3|\bx-\bX_i|^2-2(\rho_i^2+|\bw|^2)))}
{(|\bx-\bX_i|^2+\rho_i^2+|\bw|^2))^6}\ ,
\end{align}
for $i=1,2$.

Then, the first term in \eqref{HU1exp} is evaluated as
\begin{align}
\int d^3 x dz\,
(\del_M\Box\log f_1)^2&=
-\int d^3 x dz\,
(\Box\log f_1)(\Box\Box\log f_1)\nn\\
&=
\frac{1}{\rho_1^2+|\bw|^2}
\int_0^\infty du\,2\pi^2 u^3 \frac{4(u^2+2)}{(u^2+1)^2}\frac{96}{(u^2+1)^4}\nn\\
&=
\frac{1}{\rho_1^2+|\bw|^2}\frac{256\pi^2}{5}\nn\\
&\simeq
\frac{256\pi^2}{5}\left(
\frac{1}{\rho_1^2}-\frac{|\bw|^2}{\rho_1^4}+\cO(|\br|^{-4})
\right) \ ,
\label{f1f1}
\end{align}
where we have used
\begin{eqnarray}
\bu\equiv\frac{\bx-\bX_1}{\sqrt{\rho_1^2+|\bw|^2}}\ ,
\label{u1}
\end{eqnarray}
and $u\equiv |\bu|$. A similar formula for $f_2$ is obtained by replacing
$\rho_1$ with $\rho_2$ in \eqref{f1f1}.

The third term in \eqref{HU1exp} is evaluated as
\begin{align}
\int d^3 x dz\,
(\del_M\Box\log f_1)(\del_M\Box\log f_2)&=
-\int d^3 x dz\,
(\Box\log f_2)(\Box\Box\log f_1)\nn\\
&=\frac{1}{|\br|^2}
\int d^4u\,
 \frac{4(|V_1\bu+\wh\br|^2+2V_2^2)}{(|V_1\bu+\wh\br|^2+V_2^2)^2}\frac{96}{(u^2+1)^4}\ .
\label{f1f2}
\end{align}
where $\br=\bX_1-\bX_2$ and we have defined
\begin{eqnarray}
\wh\br=\frac{\br}{|\br|}\ ,~~
V_i=\frac{\sqrt{\rho_i^2+|\bw|^2}}{|\br|}\ .~~(i=1,2)
\end{eqnarray}
To evaluate the leading term in the $1/|\br|$ expansion,
we consider the limit $V_i\ra 0$. Although the integrand of \eqref{f1f2}
is divergent at $\bu=-\wh\br/V_1$ when $V_2=0$, the integral around
$\bu=-\wh\br/V_1$ is convergent. Besides, there is a suppression factor
$1/(u^2+1)^4$ that makes the contribution around $\bu=-\wh\br/V_1$
in the integral vanish in the $V_1\ra 0$ limit. Therefore, we can safely
take the $V_i\ra 0$ limit and using
\begin{align}
\int d^4u\,
\frac{4\cdot 96}{(u^2+1)^4}=64\pi^2 \ ,
\end{align}
we obtain
\begin{align}
\int d^3 x dz\,
(\del_M\Box\log f_1)(\del_M\Box\log f_2)&=
\frac{64\pi^2}{|\br|^2}+\cO(|\br|^{-4})\ .
\label{f1f2-2}
\end{align}

The last term in \eqref{HU1exp} is given by
\begin{align}
&\int d^3 x dz\,
(\del_M\Box\log f_1)(\del_M\Box\log f_3)\nn\\
&=
-\int d^3 x dz\,
(\Box\Box\Box\log f_1)\log f_3\nn\\
&=\frac{1}{\rho_1^2+|\bw|^2}
\int d^4u\,\frac{1536(3u^2-2)}{(u^2+1)^6}\log f_3\ ,
\label{f1f3}
\end{align}
with
\begin{eqnarray}
f_3=
1-\frac{\left((\by_1\cdot\by_2)-2V_1|\br|(\bw\cdot \bu)\right)^2}
{|\br|^4V_1^2(u^2+1)(|\wh\br+V_1\bu|^2+V_2^2)}\ .
\end{eqnarray}
Note that we have used the relation $(\bw\cdot\br)=0$, which follows from the
definition \eqref{w} with $\alpha=0$.
Again, to obtain the leading order terms in the $\cO(|\br|^{-1})$ expansion,
it is allowed to pick up the leading term in the integrand as
\begin{eqnarray}
\log f_3\simeq
-\frac{(\by_1\cdot\by_2)^2}
{|\br|^4V_1^2(u^2+1)}+\cO(|\br|^{-3})\ .
\label{logf3}
\end{eqnarray}
Using the formula
\begin{eqnarray}
 \int d^4
  u\frac{1536(3u^2-2)}{(u^2+1)^6}\frac{1}{u^2+1}=-\frac{128\pi^2}{5}\ ,
\end{eqnarray}
we obtain
\begin{align}
\int d^3 x dz\,
(\del_M\Box\log f_1)(\del_M\Box\log f_3)
&\simeq
\frac{128\pi^2}{5}\frac{1}{|\br|^2}\frac{\rho_2^2}{\rho_1^2}
(\ba_1\cdot\ba_2)^2+\cO(|\br|^{-3})\ ,
\label{f1f3-2}
\end{align}
and similarly
\begin{align}
\int d^3 x dz\,
(\del_M\Box\log f_2)(\del_M\Box\log f_3)
&\simeq
\frac{128\pi^2}{5}\frac{1}{|\br|^2}\frac{\rho_1^2}{\rho_2^2}
(\ba_1\cdot\ba_2)^2+\cO(|\br|^{-3})\ .
\label{f2f3}
\end{align}

As one can see in \eqref{logf3},
 $\log f_3$ is $\cO(|\br|^{-2})$ and hence
the second term in \eqref{HU1exp} does not contribute to the
leading $\cO(|\br|^{-2})$ terms in the potential.
Collecting \eqref{f1f1}, \eqref{f1f2-2}, \eqref{f1f3-2}, and
\eqref{f2f3}, we reproduce the potential \eqref{HU1}.

\section{Height of one-boson-exchange potential}
\label{sec:height}

In this appendix, we try to evaluate the height of the nucleon-nucleon
potential, in the one-boson-exchange approximation. Note that as shown
in \S\ref{sec:5-2} the one-boson-exchange model does not
describe correctly the short distance behavior. Thus, this appendix is
only for an illustration of what will happen in general when two
baryons are on top of each other in real space.

If the instantons are located within the distance of 
${\cal O}(1/M_{\rm KK})$, the one-boson-exchange potential is
(\ref{pot4d}). 
When the instantons are located at $Z_i\neq 0$, there is an
additional classical potential coming from the self-energy part \eqref{urhoz}, 
so in total, the inter-instanton
potential energy is
\begin{eqnarray}
  V = 
\frac{N_c}{16 \pi^2 a \lambda}
\frac{1}{ |\vec{X}_1-\vec{X}_2|^2 + (Z_1-Z_2)^2 }
+ 8\pi^2 a \lambda N_c
\left[ \frac{(Z_1)^2}{3}
+\frac{(Z_2)^2}{3}
\right].
\end{eqnarray}
This classical potential exhibits an interesting structure.
Let us find a minimum of this potential for fixed inter-baryon distance
in real space, $|\vec{X}_1-\vec{X}_2| = |\vec r|$. We employ a classical
approximation for $Z_i$, by taking the large $N_c$ limit, for
simplicity. Owing to 
the exchange symmetry $Z_1 \leftrightarrow Z_2$, the
potential is minimized at $Z_1 = -Z_2 = r_4/2$. 
Thus, the minimization problem is for the potential
\begin{eqnarray}
  V = 
\frac{N_c}{16 \pi^2a \lambda}
\frac{1}{ |\vec r|^2 + r_4^2 }
+ \frac{4\pi^2 a \lambda N_c}{3} r_4^2 \ .
\label{potZ}
\end{eqnarray}
The minimization condition is 
\begin{eqnarray}
 \frac{\partial V}{\partial r_4} = 
-\frac{N_c}{8 \pi^2a \lambda}
\frac{r_4}{ (|\vec r|^2 + r_4^2 )^2}
+ \frac{8\pi^2 a \lambda N_c}{3} r_4=0 \ .
\end{eqnarray}
This is solved with
\begin{eqnarray}
 |\vec r|^2 + r_4^2 = \frac{\sqrt{3}}
{8 \pi^2 a \lambda} \ .
\label{circle}
\end{eqnarray}
This forms a three-dimensional sphere around the origin in the four-dimensional
space. 
For a fixed $|\vec r|$, we obtain nonzero $r_4$
to minimize the classical potential. The instantons go away from $Z=0$
axis, to minimize the potential energy. This can be understood as
follows. The instantons have the overall $U(1)$ electric charge, so
they try to be away from each other. But at the same time there is 
an effect of the curved space-time, which tries to bring the instanton
toward the $Z=0$ axis. The balance of these two effects results in the
minimization at (\ref{circle}).
The minimum energy for a fixed inter-baryon distance is
\begin{eqnarray}
 V = \frac{N_c}{\sqrt{3}} -
 \frac{4\pi^2 a \lambda N_c}{3}  |\vec r|^2 \ .
\label{finalpot}
\end{eqnarray}
Thus, the inter-baryon potential
height is maximized at $|\vec r|=0$, with the height value
\begin{eqnarray}
 V_{\rm max} = \frac{N_c}{\sqrt{3}} \ .
\end{eqnarray}

The sphere (\ref{circle}) does not reach the region 
$|\vec r|^2 > \frac{\sqrt{3}} {8 \pi^2 a \lambda}$. 
In fact, in this region, 
the potential energy is minimized by $r_4=0$ in (\ref{potZ}), 
which results in the previous result (\ref{pot4d2}).
It is smoothly connected with (\ref{finalpot}) at 
$|\vec r|^2 = \frac{\sqrt{3}}{8 \pi^2 a
\lambda}$.

For the sphere (\ref{circle}) to make sense, its radius should be larger
than the classical radius of the instanton, (\ref{rhocl}). 
Unfortunately, both are of the same order, 
$\sim {\cal O}(1/\sqrt{8\pi^2 a \lambda})$, so one cannot trust this
sphere radius. However, we find an interesting picture of the
instantons, where the generic feature of the potential structure
suggests that instantons do not overlap in the spatial four dimensions
although they look overlapped in the spatial three dimensions.

\section{One-boson-exchange potential revisited}
\label{EPr}

In this appendix, we rederive the one-boson-exchange potential
of \S\ref{sec:5}
by summing up an infinite number of one-boson-exchange diagrams
explicitly, in the standard field-theoretical computation. 
A key ingredient for this computation is the nucleon-nucleon-meson 
cubic couplings obtained in Ref.~\citen{HSS}.\footnote{
In Ref.~\citen{HSS}, the cubic couplings
involving excited baryons with $I=J=1/2$ are also calculated.
The extension of the computation of the nucleon-nucleon potential
in this appendix to such excited states is straightforward.
}
It is found that the results are in total agreement with those
derived in \S\ref{sec:5}.
See \S 4.3 of Ref.~\citen{HSS} for the definition of the
couplings and the effective Lagrangian that we use
to obtain the Feynman rule.

\subsection{Pion exchange}
The Feynman rule for the Yukawa coupling
among a pion, nucleon $N$ and nucleon $N$ reads
\begin{align}
ig_{\pi NN}\,\gamma_5\,\tau^a~~(\mbox{isotriplet sector}) \ ,~~~~~
 i\wh g_{\pi NN}\,\gamma_5\,\tau^0~~(\mbox{isosinglet sector}) \ .
\end{align}
Consider the scattering process where two initial nucleons with 
$(p_1,s_1,I_1)$ and $(p_2,s_2,$ $I_2)$ scatter to 
the final state composed of the two nucleons labeled as
$(p'_1,s'_1,I'_1)$ and $(p'_2,s'_2,I'_2)$
by exchanging a single pion. 
Here, $p_{1},p_{2},p'_{1}$, and $p'_{2}$ are the on-shell momenta with the
nucleon mass given by $m_B$.
$s_1,s_2,s'_1$, and $s'_2$ specify the third components of the
spin of the nucleons, and
$I_{1},I_{2},I'_{1}$, and $I'_{2}$ stand for the third
components of the isospin.
It turns out that the scattering amplitudes due to the isotriplet
and isosinglet pseudo-scalar meson exchange are given by
\begin{align}
\cM^{SU(2)}_\pi=&\left(ig_{\pi NN}\right)^2\,
\sqrt{2E_1}\sqrt{2E_2}\sqrt{2E'_1}\sqrt{2E'_2}\,
\tau^a_{I'_1I_1}\tau^a_{I'_2I_2}\,
\nn\\
&~~~~~~~~~\times\ol u(p'_1,s'_1)\gamma_5 u(p_1,s_1)\,
\frac{1}{k^2+m_\pi^2}\,
\ol u(p'_2,s'_2)\gamma_5 u(p_2,s_2) \ ,
\nn\\
\cM^{U(1)}_\pi=&\cM^{SU(2)}_\pi
\Big|_{g_{\pi NN}\to \wh g_{\pi NN},\,\tau^a\to\tau^0 \,} \ ,
\end{align}
respectively. Here, $k=p_1-p'_1=p'_2-p_2$ and
\begin{align}
 \tau^a_{I'I}=\chi^{(I')\dagger}\,\tau^a\,\chi^{(I)} \ ,
\end{align}
with $\chi^{(I=1/2)}=(1,0)^T$ and $\chi^{(I=-1/2)}=(0,1)^T$
being the isospin wavefunctions.
The same notation will be used for the spin matrices 
$\sigma^a_{s's}$.
For the definition of the Dirac spinors, see Appendix B.2 in
Ref.~\citen{HSS}.
Note also that we regard the pion as being massive with $m_{\pi}\ne 0$
for the moment although the pion is massless in our model.

In the large $N_c$ and large $\lambda$ limit, the nucleon mass
$m_B$ scales as $\cO(\lambda N_c)$ so that the nonrelativistic approximation
is valid by considering the momenta to be of order one. Then
\begin{align}
& E_1=E'_1=E_2=E'_2\simeq m_B \ ,
\nn\\
&k^0\simeq\frac{1}{2m_B}(\vp_1^{\,2}-\vp_1^{\,\prime 2})=\cO(m_B^{-1}) \ ,~~~
k^2\simeq \vk^2 \ .
\end{align}
{}Furthermore, 
it can be shown that
\begin{align}
 \ol u(p'_1,s'_1)\gamma_5 u(p_1,s_1)\simeq\frac{1}{2m_B}\,
(p_1-p'_1)_a\,\sigma^a_{s'_1s_1} \ .
\end{align}
Hence,
\begin{align}
 \cM^{SU(2)}_\pi\simeq +(2m_B)^2\,\frac{g_{\pi NN}^2}{(2m_B)^2}\,
\left(\vec{\tau}_1\cdot\vec{\tau}_2\right)
\left(\vec{k}\cdot\vec{\sigma}_1\right)
\left(\vec{k}\cdot\vec{\sigma}_2\right)
\frac{1}{\vec{k}^2+m_{\pi}^2}\ ,
\label{Mpi}
\end{align}
and a similar expression holds for the $U(1)$ part.
Here, in abbreviation,
\begin{align}
 \vec{\tau}_i=\vec{\tau}_{I'_iI_i} \ ,~~~
 \vec{\sigma}_i=\vec{\sigma}_{s'_is_i} \ .~~~
(i=1,2)
\end{align}
To estimate the order of the amplitudes in
$\lambda$ and $N_c$, recall that the Yukawa couplings
are given in Ref.~\citen{HSS} as
\begin{align}
&\wh g_{\pi NN}=\frac{m_B}{f_\pi}\frac{N_c}{16\pi^3\kappa}
\vev{\frac{1}{k(Z)}}
\ ,
\quad
g_{\pi NN}=\frac{m_B}{f_\pi}\frac{16\pi\kappa}{3}
 \vev{\frac{\rho^2}{k(Z)}}
\ ,
\label{piBB}
\end{align}
with $f_\pi^2=(4/\pi)\kappa$.
Here, $\vev{~~}$ denotes the expectation value with respect to
the nucleon wavefunction given in \S\ref{sec:2}.
This implies that
\begin{align}
 g_{\pi NN}=\cO(\lambda^{1/2}\,N_c^{3/2}) \ ,~~~
 \wh g_{\pi NN}=\cO(\lambda^{-1/2}\,N_c^{1/2}) \ ,~~~
\end{align}
showing that the $SU(2)$ part dominates the $U(1)$ part.

The effective action of the nucleons is defined to reproduce
the amplitude computed above. In particular, the effective potential
should be equated with
\begin{align}
 -\wt V_{\pi}=+\frac{g_{\pi NN}^2}{(2m_B)^2}\,
\left(\vec{\tau}_1\cdot\vec{\tau}_2\right)
\left(\vec{k}\cdot\vec{\sigma}_1\right)
\left(\vec{k}\cdot\vec{\sigma}_2\right)
\frac{1}{\vec{k}^2+m_{\pi}^2}\ .
\end{align}
Note that the overall factor $(2m_B)^2$ is removed from 
(\ref{Mpi}) because this comes from the wavefunctions assigned
to the four external lines.
By Fourier-transforming this, we obtain
\begin{align}
 V_\pi(\vx)=\int \frac{d^3k}{(2\pi)^3}
  e^{i\vec k\cdot\vec x} \wt V_{\pi}\ .
\end{align}
Using the formula
\begin{align}
 \int \frac{d^3k}{(2\pi)^3}\,e^{i\vk\cdot\vx}\,\frac{1}{\vk^2+m^2}
&=\frac{1}{4\pi}\,\frac{e^{-mr}}{r} \ ,
\end{align}
with $r=|\vx|$, and also for any function $g(r)$,
\begin{align}
\left(\vec{\sigma_1}\cdot\vec{\nabla}\right)
\left(\vec{\sigma_2}\cdot\vec{\nabla}\right)g(r)
=\frac{1}{3}\left(\vec{\sigma_1}\cdot\vec{\sigma}_2\right)
\vec{\nabla}^2g(r)
+\frac{1}{3}S_{12}\left(\partial_r^2g-\frac{1}{r}\partial_rg\right)
\ ,
\end{align}
with 
\begin{align}
S_{12}=3\frac{
\left(\vec{\sigma_1}\cdot\vec{r}\right)
\left(\vec{\sigma_2}\cdot\vec{r}\right)}{r^2}
-\left(\vec{\sigma_1}\cdot\vec{\sigma_2}\right) 
\end{align} 
being the tensor operator, we find 
\begin{align}
 V_{\pi}(\vx)=\frac{g_{\pi NN}^2}{4\pi}\frac{1}{(2m_B)^2}
\left(\vec{\tau}_1\cdot\vec{\tau}_2\right)
\left[
S_{12}\frac{e^{-m_\pi r}}{r}\,
\left(\frac{m_\pi^2}{3}\!+\!\frac{m_\pi}{r}\!+\!\frac{1}{r^2}\right)
+\frac{m_{\pi}^2}{3}
\left(\vec{\sigma}_1\cdot\vec{\sigma}_2\right)
\frac{e^{-m_\pi r}}{r}
\right] \ .
\label{OPEP2}
\end{align}
Using the Goldberger-Treiman relation
\begin{eqnarray}
g_A=\frac{f_\pi g_{\pi NN}}{m_B}\ ,
\end{eqnarray}
we find that the one-pion-exchange potential
\eqref{OPEP2} agrees with the expression \eqref{OPEP}
used in \S\ref{sec:5}.
As discussed in \S\ref{sec:5}, the central force vanishes
when $m_\pi=0$ and the potential \eqref{OPEP2}
reproduces \eqref{Vlarge}, and the $n=0$ component of
the tensor force \eqref{VT}.
{}For $m_\pi \ne 0$, it is standard 
in the literature to define the coupling 
\begin{align}
 f^2=\frac{g_{\pi NN}^2}{4\pi}\left(\frac{m_{\pi}}{2m_B}\right)^2 \ ,
\end{align}
with which
\begin{align}
 V_{\pi}(\vx)=
m_\pi\,\frac{f^2}{3}(\vec\tau_1\cdot\vec\tau_2)
\left[
S_{12}
\left(1+\frac{3}{m_\pi r}+\frac{3}{m_\pi^2r^2}\right)
+\left(\vec{\sigma}_1\cdot\vec{\sigma}_2\right)
\right]\frac{e^{-m_\pi r}}{m_\pi r}
\ .
\end{align}

\subsection{Axial-vector meson exchange}

The Feynman rule for the nucleon-nucleon-axial-vector-meson cubic
couplings is
\begin{align}
g_{a^nNN}\,i\gamma_5\gamma^\mu\,\frac{\tau^a}{2} \ ~~
(\mbox{isotriplet sector})  \ ,~~~~~~
 \wh g_{a^nNN}\,i\gamma_5\gamma^\mu\,\frac{1}{2} \ ~~
(\mbox{isosinglet sector})  \ .
\end{align}
Here, $a^n\,(n=1,2,\cdots)$ is the axial-vector meson 
associated with the wavefunction $\psi_{2n}$ with the mass squared given 
by $\lambda_{2n}$.
The propagator for a massive (axial-)vector boson of mass $m$ 
is given by 
\begin{align}
 \frac{1}{k^2+m^2}\left(\eta_{\mu\nu}
+\frac{k_\mu k_\nu}{m^2}\right) \ .
\end{align}
{}From these, the amplitude of the two nucleons exchanging
an isotriplet axial-vector meson, summed over the 
species of the exchanged
mesons, becomes
\begin{align}
\cM^{SU(2)}_a=&
\sqrt{2E_1}\sqrt{2E_2}\sqrt{2E'_1}\sqrt{2E'_2}\,\,
\frac{1}{4}\left(\vec{\tau}_1\cdot\vec{\tau}_2\right)
\nn\\
&\hspace{-14mm}
\times\sum_{n\ge 1}\frac{-g_{a^nNN}^2}{k^2+\lambda_{2n}}
\left(\eta_{\mu\nu}+\frac{k_\mu k_\nu}{\lambda_{2n}}\right)
\left(\ol u(p'_1,s'_1)\gamma_5\gamma^\mu u(p_1,s_1)\right)
\left(\ol u(p'_2,s'_2)\gamma_5\gamma^\nu u(p_2,s_2)\right)
\ ,
\end{align}
and we obtain a similar expression for the isosinglet case.
Note that the cubic couplings are computed in Ref.~\citen{HSS} as
\begin{align}
&\wh g_{a^nNN}
=\frac{N_c}{32\pi^2\kappa}\langle\partial_Z\psi_{2n}(Z)\rangle \ ,
\quad
g_{a^nNN}=\frac{8\pi^2\kappa}{3}\,
\langle\rho^2\rangle\langle\partial_Z\psi_{2n}(Z)\rangle 
\ .
\label{ganBB}
\end{align}
This shows that
\begin{align}
 g_{a^nNN}=\cO(\lambda^{-1/2}N_c^{1/2}) \ ,~~~~
\wh g_{a^nNN}=\cO(\lambda^{-1/2}N_c^{-1/2}) \ ,
\end{align}
and therefore the isosinglet sector is negligible compared with
the isotriplet sector, as in the pion exchange case.

In the nonrelativistic limit, where
\begin{align}
\ol u(p',s')\gamma_5\gamma^0 u(p,s)=\cO(m_B^{-1}) \ ,
~~~
\ol u(p',s')\gamma_5\gamma^j u(p,s)=
i\sigma^j_{s's}+\cO(m_B^{-2}) \ ,
\end{align}
the scattering amplitude is dominated by
the spatial component of the axial-vector fields so that
\begin{align}
\cM^{SU(2)}_a\simeq&
(2m_B)^2\,\,
\frac{1}{4}\left(\vec{\tau}_1\cdot\vec{\tau}_2\right)
\sum_{n\ge 1}\frac{+g_{a^nNN}^2}{\vk^2+\lambda_{2n}}
\left[
\left(\vec{\sigma}_1\cdot\vec{\sigma}_2\right)
+\frac{1}{\lambda_{2n}}\,
\left(\vec{k}\cdot\vec{\sigma}_1\right)
\left(\vec{k}\cdot\vec{\sigma}_2\right)
\right]
\nn\\
=&-(2m_B)^2\,\wt V_a^{SU(2)}
\ .
\end{align}
This yields the effective
potential due to the axial-vector-meson exchange:
\begin{align}
 V_a^{SU(2)}(\vx)
%
&=-\frac{1}{4}\left(\vec{\tau}_1\cdot\vec{\tau}_2\right)
\sum_{n\ge 1}g_{a^nNN}^2\!\!
\int \!\!\frac{d^3k}{(2\pi)^3}\,e^{i\vec k\cdot\vec x}
\left[
\frac{\left(\vec{\sigma}_1\cdot\vec{\sigma}_2\right)}
{\vk^2+\lambda_{2n}}
+\frac{1}{\lambda_{2n}}\,
\frac{
\left(\vec{k}\cdot\vec{\sigma}_1\right)
\left(\vec{k}\cdot\vec{\sigma}_2\right)}
{\vk^2+\lambda_{2n}}
\right]
\nn\\
&=-\frac{1}{4}\left(\vec{\tau}_1\cdot\vec{\tau}_2\right)
\sum_{n\ge 1}g_{a^nNN}^2
\left[
S_{12}\,\frac{1}{\lambda_{2n}\,r^2}
\left(
1+\sqrt{\lambda_{2n}}\,r+\frac{1}{3}\lambda_{2n}r^2
\right)
-\frac{2}{3}\left(\vec{\sigma}_1\cdot\vec{\sigma}_2\right)
\right]Y_{2n}(r) 
\nn\\
&=V_{{\rm C}}^{(a)}+S_{12}V_{{\rm T}}^{(a)} \ ,
\end{align}
where
\begin{align}
V_{{\rm C}}^{(a)}&=
\frac{1}{6}\left(\vec{\tau}_1\cdot\vec{\tau}_2\right)
\left(\vec{\sigma}_1\cdot\vec{\sigma}_2\right)
\sum_{n\ge 1}g_{a^nNN}^2
Y_{2n}(r)\ , \\
V_{{\rm T}}^{(a)}&=
-\frac{1}{12}\left(\vec{\tau}_1\cdot\vec{\tau}_2\right)
\sum_{n\ge 1}g_{a^nNN}^2
\,\frac{1}{\lambda_{2n}}
\left(
\frac{3}{r^2}+\frac{3\sqrt{\lambda_{2n}}}{r}
+\lambda_{2n}
\right)
Y_{2n}(r) \ .
\end{align}
Using (\ref{ganBB}),  it is easy to show that this
agrees with the $n=2,4,6,\cdots$ components of
(\ref{VC}) and (\ref{VT}).

\subsection{Vector meson exchange}

The Feynman rule states that
for the nucleon-nucleon-$v^n$ cubic couplings, we assign
\begin{align}
 i\,\frac{\tau^a}{2}\left(
g_{v^nNN}\,\gamma^\mu-\frac{h_{v^nNN}}{2m_B}\,
\sigma^{\mu\nu}k_\nu\right) \ ,~~~~
 i\,\frac{1}{2}\left(
\wh g_{v^nNN}\,\gamma^\mu-\frac{\wh h_{v^nNN}}{2m_B}\,
\sigma^{\mu\nu}k_\nu\right) \ ,
\end{align} 
for the isotriplet and isosinglet cases, respectively. 
Here, $v^n\,(n=1,2,\cdots)$ is the vector meson 
associated with the wavefunction $\psi_{2n-1}$, whose mass squared is
equal to $\lambda_{2n-1}$,
and $k$ is the momentum of the vector meson flowing
outwards from
the vertex.
It follows that the two-nucleon scattering amplitude due to
the exchange of an isotriplet vector meson, summed over the infinite
tower of the vector meson species, 
is given by
\begin{align}
 \cM_v^{SU(2)}&=\sqrt{2E_1}\sqrt{2E_2}\sqrt{2E'_1}\sqrt{2E'_2}\,\,\,
\frac{-1}{4}\,\left(\vec{\tau}_1\cdot\vec{\tau}_2\right)
\nn\\
&\hspace{-10mm}
\times\sum_{n\ge 1}\bigg[
\ol u(p'_1,s'_1)\left(g_{v^nNN}\,\gamma^\mu-\frac{h_{v^nNN}}{2m_B}\,
\sigma^{\mu\rho}k_\rho\right) u(p_1,s_1)
\,\,
 \frac{1}{k^2+\lambda_{2n-1}}\left(\eta_{\mu\nu}
+\frac{k_\mu k_\nu}{\lambda_{2n-1}}\right) 
\nn\\
&\qquad\quad
\times\ol u(p'_2,s'_2)\left(g_{v^nNN}\,\gamma^\nu+\frac{h_{v^nNN}}{2m_B}\,
\sigma^{\nu\sigma}k_\sigma\right) u(p_2,s_2)
\bigg] \ .
\end{align}
As before, a similar expression follows for the $U(1)$ part.

In the nonrelativistic limit, we have
\begin{align}
 \ol u(p',s')\gamma^0\,u(p,s)&=-i\delta_{s's}+\cO(m_B^{-2}) \ ,
\nn\\
 \ol u(p',s')\gamma^j\,u(p,s)&=-\frac{i}{2m_B}\Big[
(p+p')_j\,\delta_{ss'}+i\epsilon_{jla}(p-p')_l\,\sigma^a_{s's}
\Big]
+\cO(m_B^{-2}) \ ,
\nn\\
 \ol u(p',s')\sigma^{0j}\,u(p,s)&=-\frac{i}{2m_B}\Big[
(p-p')_j\,\delta_{ss'}+i\epsilon_{jla}(p+p')_l\,\sigma^a_{s's}
\Big]
+\cO(m_B^{-2}) \ ,
\nn\\
 \ol u(p',s')\sigma^{jk}\,u(p,s)&=-\epsilon^{jka}\,\sigma^a_{s's}
+\cO(m_B^{-2}) \ .
\end{align}
{}Furthermore, we note that the cubic coupling constants
obtained in Ref.~\citen{HSS} are given by
\begin{align}
&g_{v^nNN}=\vev{\psi_{2n-1}(Z)}  \ ,~~~
h_{v^nNN}=\frac{16\pi^2\kappa m_B}{3}\,\vev{\rho^2}\vev{\psi_{2n-1}(Z)} \ ,
\nn\\
&\wh g_{v^nNN}=N_c\vev{\psi_{2n-1}(Z)}  \ ,~~~
\wh h_{v^nNN}=N_c\left(\frac{m_B}{16\pi^2\kappa}-1\right)\,
\vev{\psi_{2n-1}(Z)} \ ,
\end{align}
which imply
\begin{align}
&g_{v^nNN}=\cO(\lambda^{-1/2} N_c^{-1/2})\ ,~~
h_{v^nNN}=\cO(\lambda^{1/2} N_c^{3/2})\ ,\nn\\
&\wh g_{v^nNN}=\cO(\lambda^{-1/2} N_c^{1/2})\ ,~~
\wh h_{v^nNN}=\cO(\lambda^{-1/2} N_c^{1/2})\ .
\end{align}
Then, among the spinor bilinear forms appearing
in the amplitudes, the leading ones for large $\lambda$
and large $N_c$ are
\begin{align}
 \ol u(p',s')\left(g_{v^nNN}\,\gamma^j-\frac{h_{v^nNN}}{2m_B}\,
\sigma^{j\rho}(p-p')_\rho\right) u(p,s)
&\simeq \frac{h_{v^nNN}}{2m_B}\epsilon_{jla}(p-p')_l\sigma^a_{s's}
+\cdots \ ,
\\
 \ol u(p',s')\left(\wh g_{v^nNN}\,\gamma^0-\frac{\wh h_{v^nNN}}{2m_B}\,
\sigma^{0\rho}(p-p')_\rho\right) u(p,s)
&\simeq -i\wh g_{v^nNN}\,\delta_{s's}+\cdots \ .
\end{align}
This shows that for the isotriplet vector mesons, the spatial components 
dominate the amplitude, while for the isosinglet vector mesons,
the time component does.
Consequently,
\begin{align}
 \cM^{SU(2)}_v&\simeq (2m_B)^2\,\frac{1}{4}\,
\left(\vec{\tau}_1\cdot\vec{\tau}_2\right)
\epsilon_{jla}\,k_l\,\sigma^a_{s'_1s_1}\,
\epsilon_{jmb}\,k_m\,\sigma^b_{s'_2s_2}
\sum_{n\ge 1}\frac{1}{\vk^2+\lambda_{2n-1}}
\left(\frac{h_{v^nNN}}{2m_B}\right)^2
\nn\\
&=-(2m_B)^2\wt V_v^{SU(2)} \ ,
\\
 \cM^{U(1)}_v&\simeq -(2m_B)^2\,\frac{1}{4}\,
\sum_{n\ge 1}\frac{\wh g_{v^nNN}^2}{\vk^2+\lambda_{2n-1}}
=-(2m_B)^2\wt V_v^{U(1)} \ .
\end{align}
Fourier-transforming the effective potentials gives
\begin{align}
 V_v(\vx)=\int\frac{d^3k}{(2\pi)^3}\,
 e^{i\vec k\cdot\vec x}\left(\wt V^{SU(2)}+\wt V^{U(1)}
\right)
=V_{{\rm C}}^{(v)}(\vx)+S_{12}V_{{\rm T}}^{(v)}(\vx) \ ,
\end{align}
with
\begin{eqnarray}
&&V_{{\rm C}}^{(v)}(\vx)=
-\frac{1}{6}\,
\left(\vec{\tau}_1\cdot\vec{\tau}_2\right)
\left(\vec{\sigma}_1\cdot\vec{\sigma}_2\right)
\!\sum_{n\ge 1}\!\lambda_{2n-1}\!\left(
\!\frac{h_{v^nNN}}{2m_B}\!\right)^2
\!\!
Y_{2n-1}(r)
-\!\frac{1}{4}\!\,\sum_{n\ge 1}\wh g_{v^nNN}^2\,Y_{2n-1}(r) \ ,
\nn\\
\\ 
&&V_{{\rm T}}^{(v)}(\vx)=
 \frac{1}{12}\, \left(\vec{\tau}_1\cdot\vec{\tau}_2\right)
\sum_{n\ge 1}\left(\frac{h_{v^nNN}}{2m_B}\right)^2\,
\left(\lambda_{2n-1}+\frac{3\sqrt{\lambda_{2n-1}}}{r}
+\frac{3}{r^2}\right)Y_{2n-1}(r) \ .
\end{eqnarray}
Again, this is in agreement with the $n=1,3,5,\cdots$
components of \eqref{VC} and \eqref{VT}.

\vspace*{1.5cm}

\noindent
{\bf Note added :}
When preparing this paper, we became aware of 
Ref.~\citen{Kim:2009iy}, which overlaps partly with our strategy. 

\noindent
{\bf Note added in the second version:}
The added appendix \ref{EPr} has some overlaps
with Ref.~\citen{Kim:2009sr}, which appeared when we were preparing
the revised version.

\end{document}